# Displacement-Field–Induced Ionization of Defect States in High-Quality hBN

Ilia Begichev[1,2], Daria Belotcerkovtceva[1,2], Iakov Reznikov[1,2], Sergey Slizovsky[3,4], Weijie Lu[1,2], Nansi Zhou[1,2], Ian Babich[1,2], Makars Šiškins[2,5], Kostya S. Novoselov[1,2], Vladimir I. Fal'ko[3,4], Kenji Watanabe[6], Takashi Taniguchi[6], Alexey I. Berdyugin[1,2,7]

[1]Department of Materials Science and Engineering, National University of Singapore, Singapore 117575, Singapore.
[2]Institute for Functional Intelligent Materials, National University of Singapore, Singapore 117544, Singapore
[3]Department of Physics and Astronomy, University of Manchester, Manchester M13 9PL, UK
[4]National Graphene Institute, University of Manchester, Manchester M13 9PL, UK
[5]School of Physics and Astronomy, University of Southampton, Southampton SO17 1BJ, UK
[6]National Institute for Materials Science, Tsukuba 305-0044, Japan
[7]Department of Physics, National University of Singapore, Singapore 117551, Singapore.

**Encapsulation of two-dimensional (2D) materials with hexagonal boron nitride (hBN) is widely used to fabricate ultraclean van der Waals (vdW) heterostructures for optoelectronic and quantum transport studies. These devices are often operated under strong out-of-plane displacement fields that tune the electronic properties of the active 2D material, while the surrounding hBN is typically assumed to remain electronically inert. This assumption, however, neglects defect states that are present even in high-quality hBN crystals. Here, we probe their response to the applied displacement field using monolayer graphene encapsulated between hBN crystals as a sensitive charge sensor. Increasing the displacement field induces notable charge transfer between graphene and hBN, producing additional electron doping of $\approx 10-20\times 10^{9}\ \mathrm{cm}^{-2}$ even at the moderate displacement field ($\sim 0.5\ \mathrm{V\,nm}^{-1}$), while simultaneously increasing carrier-density inhomogeneity of the graphene layer. Both effects are consistent with ionization of defect states within the hBN bandgap, with the bulk density $\sim 1-7\times 10^{-5}\ \mathrm{nm}^{-3}$ varying across different hBN crystals. Our results show that displacement-field-induced defect ionization in hBN can modify the electronic response of vdW heterostructures, with direct implications for device design and interpretation of experiments.**

Encapsulation with hexagonal boron nitride (hBN) has become a standard technique for the fabrication of high-quality van der Waals heterostructures for both quantum transport and photonic studies[1–4]. In such devices, the active material of interest is typically encapsulated with relatively thick ($10-50\ \mathrm{nm}$) dielectric hBN layers that serve as atomically smooth substrates, protecting the material from the environment. These days, such heterostructures are often studied under substantial out-of-plane electric displacement fields $D$ up to $1-2\ \mathrm{V/nm}$[2,5–13], which are used to tune the band structure and corresponding electronic properties of encapsulated active materials. It is commonly assumed that the applied electric field does not affect encapsulating hBN crystals, which are expected to behave as ideal dielectric layers. This assumption is largely justified by the wide bandgap of hBN ($\sim 6\ \mathrm{eV}$), which suppresses hybridization between hBN valence or conduction bands and the electronic states responsible for transport in the active material. Consequently, any electric-field-induced modifications of the hBN band structure are generally considered irrelevant for the electronic response of the heterostructure.

However, this simplified picture neglects the presence of impurities and structural defects, which are known to exist even in the highest-quality hBN crystals and give rise to deep mid-gap states[14–36]. These defect states can interact with adjacent conducting layers, leading to charge transfer[17–27,35,36] and additional scattering[21,29,30]. Moreover, several earlier studies reported gate hysteresis under applied $D$ in bilayer graphene aligned with hBN heterostructures[5,8] and attributed it to exotic electronic ferroelectricity. However, similar hysteresis was later observed in systems without an hBN-induced moiré potential, including single-layer graphene and multilayer twisted devices[6,10–13]. Some of the later works suggest that, in certain cases, such hysteresis may originate from defects in hBN crystals, highlighting the need for further characterization of hBN under an applied displacement field.

Despite the increasing recognition of the importance of hBN defect states, their response to an applied displacement field remains challenging to isolate experimentally. In most heterostructures, $D$ directly modifies the electronic band structure of the active material, producing a strong and non-trivial field response that dominates the optoelectronic behavior of the device.

In this work, we overcome this limitation by using monolayer graphene to characterize the effect of displacement field on hBN impurities in high-quality devices. Unlike most materials, the band structure of monolayer graphene is immune to the applied $D$, while its electron transport near the charge neutrality point (CNP) is highly sensitive to charged defects in the surrounding hBN medium, which are known to produce a residual charge inhomogeneity of $2–5 \times 10^9\ \mathrm{cm}^{-2}$, present even in the highest-quality hBN-encapsulated devices[1,2]. Previously, this inhomogeneity was only suppressed by suspending graphene[37,38], thereby eliminating substrate effects, or by employing proximity Coulomb screening to suppress electrostatic background from the dielectric medium[39–41]. Consequently, the electron transport near graphene CNP in high-quality devices, such as those used in this work, can be used to characterize the behavior of defect states in hBN crystals, and any changes under applied $D$ can be directly attributed to the modifications of those states.

Previous works have attempted to use monolayer graphene to characterize the impurities within the hBN and have reported shifts in graphene doping attributed to charge transfer between graphene and defect states in hBN. However, in those works, the effects were amplified by experimental conditions irrelevant to modern high-quality devices. These include the use of intentionally carbon-doped hBN crystals[21,22], neutron-irradiated hBN[27], short electric-field pulses through a $\mathrm{Si/SiO_2}$ back gate reaching up to $\sim 10\ \mathrm{V/nm}$[19,23], and intentional laser irradiation used to ionize defect states in hBN[17]. As a result, the behavior of defect states in pristine high-quality hBN crystals under applied electric fields remains largely unexplored.

**Results**

Our devices consist of monolayer graphene encapsulated between high-quality hBN crystals and with graphite top and bottom gates that enable independent control of carrier density $n$ and displacement field $D$, allowing the Fermi level in graphene to be tuned close to the CNP while applying large $D$. The structure of our devices is shown in Fig. 1a. In total, we have fabricated and tested five double-graphite-gated devices, shown further in the text, all showing similar behavior. Prior to fabrication, heterostructures were characterized by high-resolution atomic force microscopy (AFM) to find bubble-free regions, following well-established protocols for graphene device fabrication to avoid the influence of interface-confined contamination. Further device fabrication details are provided in Supplementary Information (SI) Section 1.

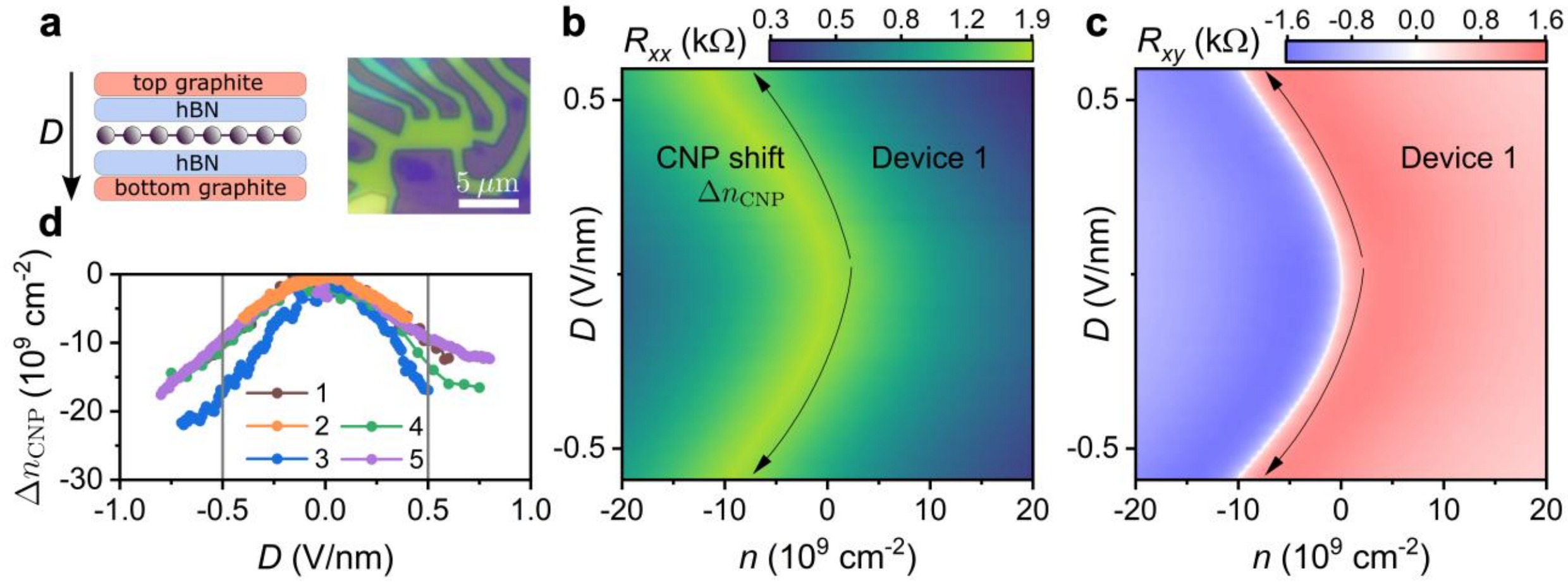

**Figure 1. Displacement-field-induced charge transfer between graphene and hBN.** a, Left: Schematic image of a double-graphite-gated device with monolayer graphene under an applied out-of-plane displacement field $D$. Right: Optical image of device 1. Scale bar, 5 $\mu$m. b, c, Longitudinal $R_{\rm xx}$ and Hall $R_{\rm xy}$ resistance maps as a function of carrier density $n$ and displacement field $D$ at $T = 100{\rm K}$ for device 1. Black arrows highlight the shift of graphene CNP as a function of $D$. d, Comparison of CNP shift as a function of $D$ for devices 1-5. Grey lines indicate $D = \pm 0.5$ V/nm, chosen for further comparison of $\Delta n_{\rm CNP}$.

To probe the effect of the displacement field $D$, we first characterized our devices at elevated temperatures of $100 - 200{\rm K}$ to suppress ballistic and mesoscopic effects near the CNP, which could otherwise complicate the interpretation of the transport data. Figure 1b,c shows a representative longitudinal $R_{\rm xx}$ and Hall $R_{\rm xy}$ resistance maps measured at zero and non-quantizing magnetic field $B$ as a function of $n$ and $D$. The most prominent effect in these maps is a shift of the graphene CNP towards negative carrier density with increasing $D$, highlighted by arrows. The same behavior is observed in all five devices, see Fig. 1d and SI Section 4. Importantly, this effect cannot be explained by inaccuracies in the gate-voltage-to-density calibration: any linear correction would merely rotate the CNP trajectory in the $n - D$ plane rather than eliminate its curvature (see SI Sections 2, 3).

To get further insights about this behavior, we have characterized the residual carrier density $n_{\rm res}$ in our devices when the Fermi level of graphene is tuned towards the CNP, and its evolution under an applied displacement field $D$. While at low temperatures residual carrier density $n_{\rm res}$ is dominated by the charge fluctuations created by the disorder $n_{\rm dis}$, at elevated temperatures $T > 50 - 100{\rm K}$ in high-quality devices, thermal excitations $n_{\rm th}$ close to CNP become comparable to the $n_{\rm dis}$, resulting in a higher residual carrier density (See SI Section 6). In this regime, the total residual density can be described as[42,43]:

$$n_{\rm res} = n_{\rm dis} + n_{\rm th} \qquad (1)$$

However, the amplitude of thermal excitations is set by the graphene band structure and therefore provides a constant background at a given temperature, which is not expected to depend on device quality or $D$. Consequently, all $D$-induced and device-specific changes $\Delta n_{\rm res}$ can be attributed to the changes of disorder $\Delta n_{\rm dis}$ within the device.

**$D$-dependence of device inhomogeneity at elevated temperatures.**

To quantify the dependence of residual carrier density $n_{\rm res}$ and corresponding $n_{\rm dis}$ on $D$, we have used two independent approaches. The first is the half-width at half-maximum (HWHM) $n_{\rm HWHM}$ of the longitudinal resistance $R_{\rm xx}$ curves[44–46], and the second is the maximum value of the Hall effect measured at a fixed small magnetic field as a function of gate voltage[47,48],

$$n_{\text{Hall}} = \frac{B}{e(R_{xy})_{\text{peak}}} \quad (2)$$

Although the latter estimation may depend on the magnetic field strength $B$ (see SI Section 7), it remains suitable for direct comparison between different values of $D$ within the same device.

Figures 2a-c show results for device 1, which has the lowest $n_{\text{dis}}$, while panels 2d-f correspond to device 5, with the highest disorder. Data for the remaining devices are presented in the SI Section 4. For both devices shown in Fig. 2, increasing the displacement field leads to the monotonic broadening of the $R_{xx}$ curves (panels a, d), indicating an increase in the inhomogeneity level. At the same time, the amplitude of the Hall curves shown in Fig. 2b,e is suppressed, and the Hall peak shifts away from the CNP (see inset in Fig. 2b), which is also consistent with increasing disorder $n_{\text{dis}}$.

Figures 2c and 2f show the behavior of $n_{\text{HWHM}}$ and $n_{\text{Hall}}$ as a function of displacement field. For both devices, $n_{\text{HWHM}}$ and $n_{\text{Hall}}$ increase monotonically with $D$ and yield similar values of inhomogeneity for a given device. While both devices exhibit a similar trend, the relative changes in device inhomogeneity under applied $D$ differ significantly. In Fig. 2c corresponding to device 1, the relative change of $n_{\text{HWHM}}$ is below $1\%$ between $D = 0$ and $0.5\ \text{V/nm}$. In contrast, the estimated inhomogeneity increases by $\sim 100\%$ within the same $D$ range for device 5 (Fig. 2f).

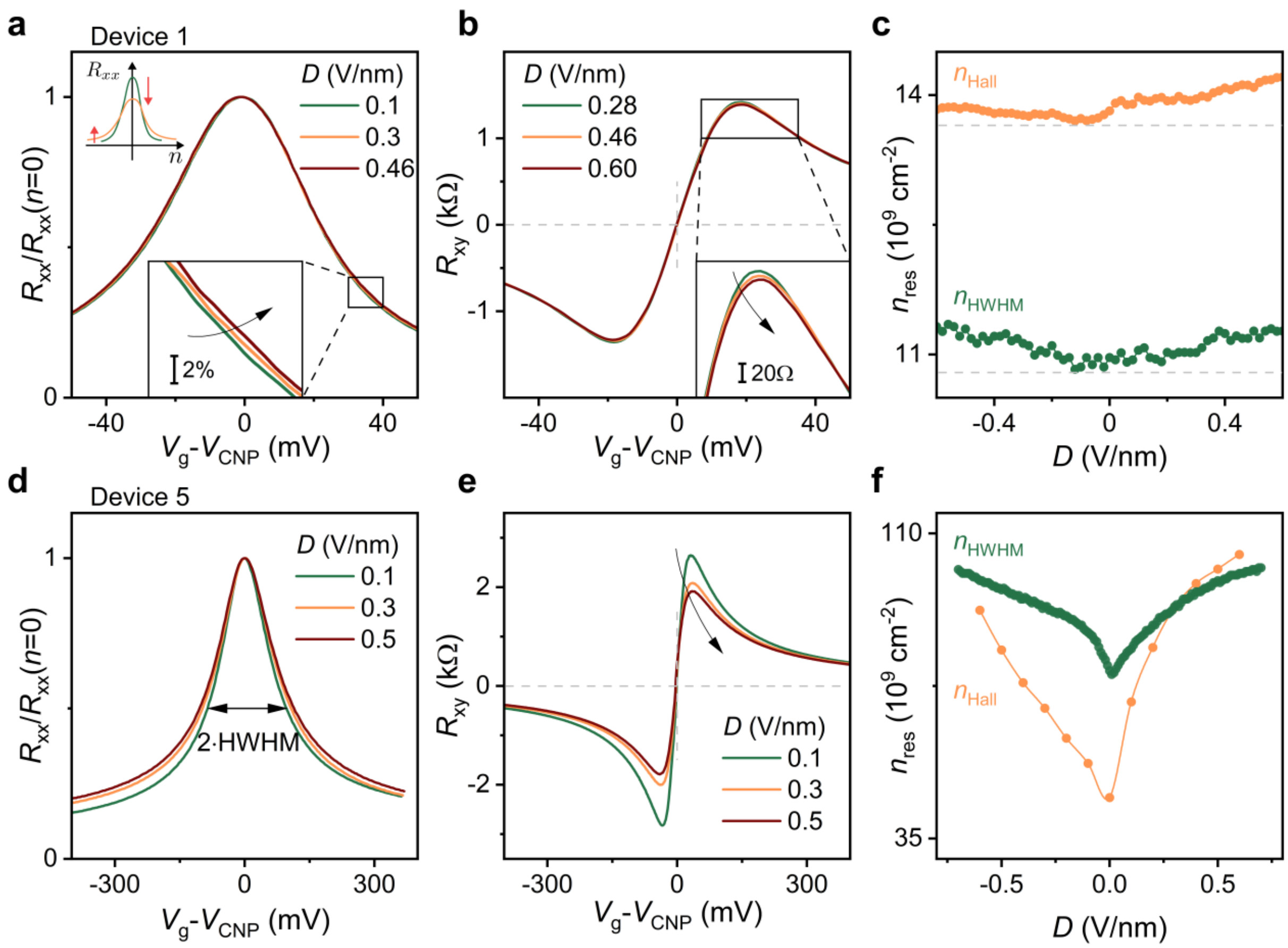

**Figure 2. Change of graphene inhomogeneity under displacement field**. a,b, Longitudinal $R_{xx}$ and Hall $R_{xy}$ resistances versus gate-voltage with respect to the Dirac point $V_g - V_{\text{CNP}}$ for representative values of $D$, measured in device 1 at $B = 0$ and $30\ \text{mT}$, respectively. Insets highlight the progressive broadening of the normalized $R_{xx}$ and suppression of $R_{xy}$ peaks under applied $D$. The schematic in (a) illustrates the evolution of $R_{xx}(V_g)$: increasing $D$ suppresses the peak while increasing the HWHM. c, The change of residual carrier density in device 1 under an applied displacement field quantified by the change of $n_{\mathrm{HWHM}}$ and $n_{\mathrm{Hall}}$ as a function of $D$. Panels (a)–(c) were measured at $T = 100\mathrm{K}$. d-f, Similar as (a)-(c), but measured in the device 5 at $T = 200\mathrm{K}$, with Hall effect measured at $B = 300\ \mathrm{mT}$.

**Low temperature characterization.**

While the change of device inhomogeneity in our highest-quality device 1 under applied $D$ is relatively small - making it barely detectable at elevated temperatures - its helium-temperature transport properties shown in Fig. 3 provide a clearer demonstration of the increased disorder.

At higher carrier densities, the device exhibits typical hallmarks of high-quality graphene, including ballistic transport signatures and a mean free path limited by the device size. Its exceptionally high electronic quality is further evidenced by the fan diagram measurements in Fig. 3a,b, where Shubnikov–de Haas oscillations emerge already at $B \approx 20–30\mathrm{mT}$. Moreover, in this device, we observe non-linear trajectories of Landau-level gaps in the fan diagram Fig. 3a and non-linear dependence of the Hall carrier density on gate voltage Fig. 3c. These effects arise from quantum-capacitance corrections that become significant near the CNP when the applied gate voltage becomes comparable to the change in Fermi energy[39,40]. Such behavior is usually difficult to resolve, as small geometrical capacitance and residual disorder mask this regime, and this observation further highlights the high electronic quality of this device.

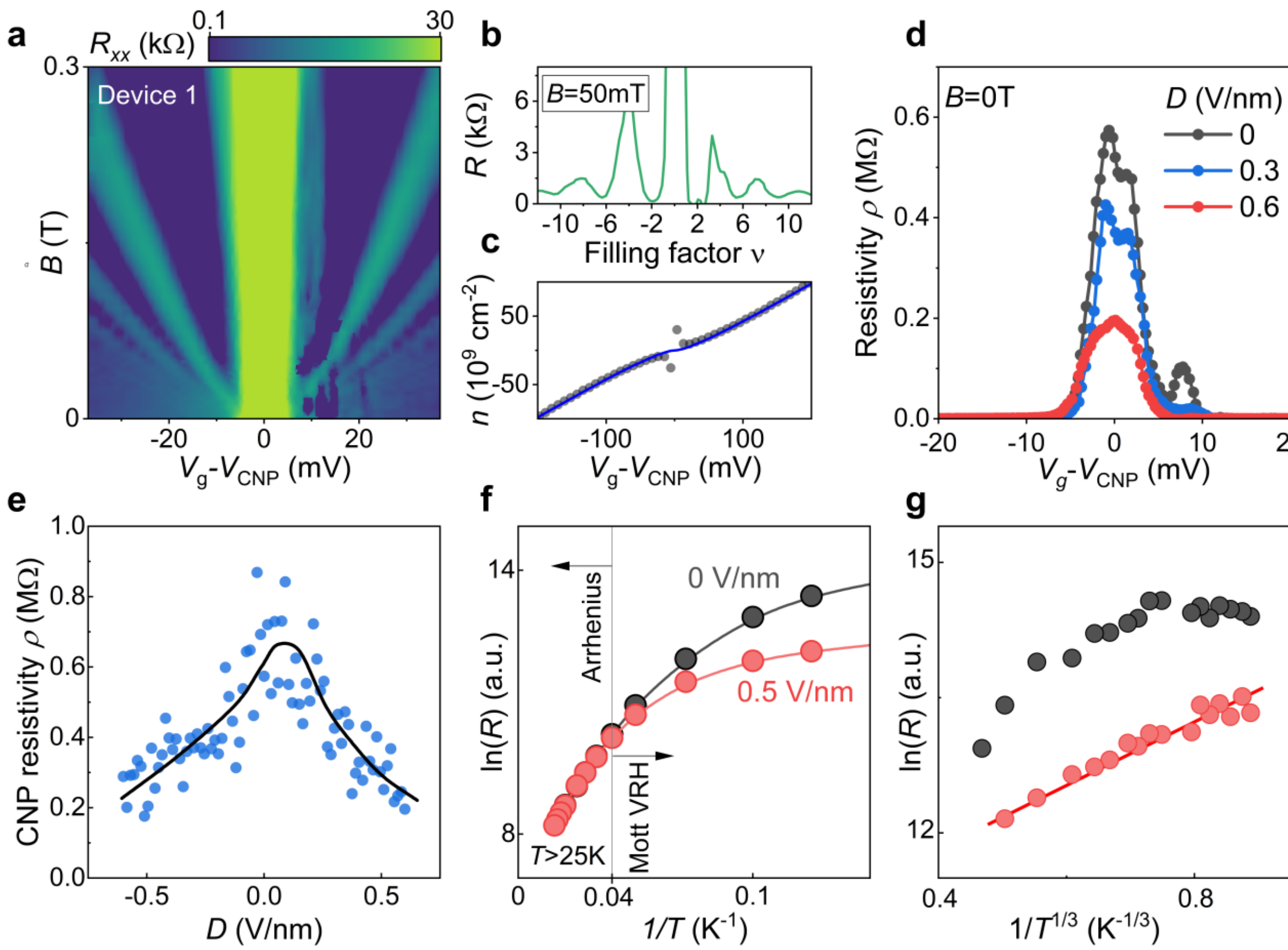

**Figure 3. Transport in the gapped regime under a displacement field.** a, Longitudinal $R_{\mathrm{xx}}$ versus gate-voltage with respect to the Dirac point $V_{\mathrm{g}} - V_{\mathrm{g0}}$ and magnetic field $B$ at $D = 0.2\ \mathrm{V/nm}$. The low-$B$ parabolic Landau fans arise from quantum-capacitance effects. b, $R_{\mathrm{xx}}$ at 50 mT versus filling factor $\nu$ obtained from the non-linear gate calibration (see SI) c, Non-linear calibration $n$ as a function of $V_{\mathrm{g}} - V_{\mathrm{CNP}}$ at 100 K: grey points are obtained from the Hall measurements at 100 mT, and the blue line is a fit which accounts for quantum capacitance. d, Resistivity at $B = 0$ for different $D$. e, Maximum CNP resistivity versus $D$; black line is a guide to the eye. f, Temperature dependence of CNP resistivity at $D = 0, 0.5\ \mathrm{V/nm}$. Above 25 K, the data are well described by an Arrhenius law, whereas at lower temperatures, they are more consistent with Mott VRH. g, Temperature dependence of the resistance below 25 K plotted versus $T^{-1/3}$, highlighting Mott VRH; the red line is a guide to the eye.

Apart from quantum-capacitance corrections, the high electronic quality of this device enables the resolution of the gap at the graphene CNP shown in Fig. 3d. At $D = 0$, the CNP resistivity of the gapped state approaches $\sim 1\mathrm{M\Omega}$, in contrast to the 2-5 $\mathrm{k\Omega}$ typically observed in lower-quality devices[49,50]. The presence of the CNP gap in our device can be attributed to inversion-symmetry breaking induced by the hBN substrate. Although gap opening is typically associated with intentional hBN–graphene alignment, several studies have also reported a small gap in nominally non-aligned devices[39,51], similar to this device, and likewise attributed it to the impact of the hBN substrate.

To probe the effect of the displacement field, we measured the gap resistivity at different $D$ shown in Fig. 3d,e. We find that CNP resistivity decreases significantly under applied displacement field from $\sim 0.7\ \mathrm{M\Omega}$ at $D = 0$ to 0.2-0.3 $\mathrm{M\Omega}$ at $D \approx 0.5\ \mathrm{V/nm}$. This behavior is characteristic of increasing disorder in an insulating state[52,53], where additional disorder enhances hopping conductivity via in-gap states.

To gain further insight, we measured the temperature dependence of CNP resistivity at $D = 0$ and $0.5\ \mathrm{V/nm}$, shown in Fig. 3f,g. At elevated temperatures $25 < T < 100\ \mathrm{K}$, CNP resistivity changes exponentially for both values of $D$, following an Arrhenius law with an activation energy of about $7\ \mathrm{meV}$. This indicates that the observed suppression of CNP resistivity in Fig. 3e is not related to a variation of the gap size. At lower temperatures $1.5 < T < 25\ \mathrm{K}$, however, the temperature dependence Fig. 3g is consistent with Mott variable-range hopping between disorder-induced localized states[52–54], $\ln R \propto T^{-1/3}$. Notably, the data at $D = 0.5\ \mathrm{V/nm}$ are significantly better described by this model than the analogous data at $D = 0\ \mathrm{V/nm}$, which can again be interpreted as a signature of increased disorder in the device under an applied displacement field.

**Discussion**

Both the charging of graphene and the increase of graphene inhomogeneity under applied displacement field, observed consistently across all devices in this work, can be attributed to displacement-field-induced ionization of defect states in hBN, as schematically illustrated in Fig. 4a. The applied electric field tilts the hBN bands, bringing in-gap defect states into resonance with graphene's Fermi level and enabling charge transfer. The resulting ionized defects act as charged impurities, enhancing electrostatic potential fluctuations in graphene and thereby increasing disorder.

Despite the substantial body of literature devoted to defects in hBN, there is still no clear consensus regarding the dominant impurity types, their exact positions within the hBN bandgap, or their exact energetic alignment relative to the graphene CNP. For example, angle-resolved photoemission spectroscopy (ARPES) studies place the graphene CNP approximately $2.5 - 2.8\ \mathrm{eV}$ above the hBN valence-band edge for epitaxial graphene and graphite on hBN[55,56]. But at the same time, photoelectron emission microscopy (PEEM) studies of monolayer graphene devices have reported significantly lower values, placing the graphene CNP only $1.6 - 2\ \mathrm{eV}$ above the hBN valence band[57].

The literature is even less conclusive about the properties of defect states. Several experimental works have demonstrated charge transfer between graphene and acceptor-like defects in hBN[18,21,22,27]. Although in some of those works, the effect was intentionally enhanced by using carbon-doped hBN crystals[21,22] with a higher number of carbon substituting nitrogen sites $\mathrm{C_N}$, or by using ion-irradiated hBN crystals[27], with an increased density of boron vacancies $\mathrm{V_B}$, these results indicate that relevant acceptor states can lie close to the graphene CNP. This picture is further supported by DFT calculations[31–33], which place those acceptor states approximately $1.4 - 2.2\ \mathrm{eV}$ above the hBN valence-band edge, i.e., within energetic proximity to the graphene CNP.

At the same time, multiple earlier studies indicate the presence of donor-like defects in hBN[17,18,23,24]. In particular, scanning tunnelling microscopy (STM) experiments on carbon-doped hBN[24] reported donor-like states attributed to carbon substituting boron sites $\mathrm{C_B}$, while photodoping experiments[17] reported substantial electron doping of graphene associated with donor-type impurities in pristine hBN crystals ionized by the light. However, in contrast to acceptors, this picture is less supported by DFT calculations[24,31–33], which place typical donor-like states such as $\mathrm{C_B}$ and nitrogen vacancies $\mathrm{V_N}$ in the range of $3.0 - 3.7\ \mathrm{eV}$ above the hBN valence-band edge, which is slightly higher than graphene CNP. This discrepancy may indicate either an overestimation of donor level energies or the involvement of more complex defect configurations. In this context, defect complexes such as $\mathrm{C_B V_N}$ can also act as donor states, with their energies lying slightly below graphene CNP, making them plausible candidates for ionization with an electric field[34–36].

A simplified summary of the discussed literature review with the relevant energy landscape is presented in Fig. 4b, which shows the positions of the graphene CNP as well as acceptor- and donor-like defect states within the hBN bandgap. The shaded ranges reflect the significant spread in reported values across different experimental and theoretical works. We note that Fig. 4b includes only those defect states located relatively close to the graphene CNP, and does not attempt to catalogue all possible defects, such as oxygen-related substitutions and others with substantially higher energies, which are unlikely to contribute under the investigated conditions.

Based on Fig. 4b, both acceptor states below CNP and donor states above it can, potentially, be in an ionized state even without applied $D$, and contribute to disorder and consequently the residual device doping. In agreement with this, we have found that at zero gate voltages, all our devices exhibit a noticeable hole doping $n_0 \sim 20 - 40 \times 10^9\ \mathrm{cm}^{-2}$ (see SI Section 2 and Supplementary Table S1), indicating the dominant charge transfer from the deep acceptor states, which is usually subtracted during data analysis to enable an easier comparison of data at different $D$.

Under applied $D$, however, we observed the opposite trend, with the displacement field activating electron doping. This can be attributed to either deionization of negatively charged acceptor states, ionization of donor-like states, or a combination of both processes. So far, we conclude that under applied $D$ both processes are likely to happen simultaneously in our devices, but that donor-state ionization dominates the net response under applied fields, while acceptor-related states primarily determine the zero-field doping.

This conclusion is further supported by observed behavior of resistivity in the gapped state in device 1. Whilst this sample shows monotonic electron doping under applied $D$ Fig. 1b,c, the CNP gap resistance Fig. 3e evolves not monotonically, but exhibits pronounced up and down fluctuations, which originate not from the electronic noise in our measurements, but rather from rapid changes of the device state as $D$ is varied (refer to SI Fig. S7 for measurement details). Similar resistance changes were also observed in Device 2, shown in SI Section 4. This behavior is consistent with the displacement field simultaneously deionizing acceptor states, thereby reducing $n_{\mathrm{dis}}$ and increasing the resistivity, and ionizing donor states, which contribute in the opposite direction. Nevertheless, the overall trend across all devices indicates that donor-state ionization is the dominant process within the studied range of displacement fields, leading to a net increase in $n_{\mathrm{dis}}$ under applied $D$ shown in Fig. 2.

To get further insights about defect states in our hBN crystals, we used both data for CNP shift $\Delta n_{\mathrm{CNP}}$ and for increasing disorder $\Delta n_{\mathrm{dis}}$ independently to extract the bulk density of impurities contributing to observed effects. The approach used to fit CNP shift gives results in the range of $\sim 1 - 7 \times 10^{-5}\ \mathrm{nm}^{-3}$ for our devices, whereas analysis based on $\Delta n_{\mathrm{dis}}$ gives significantly lower values of $\sim 0.1 - 4.0 \times 10^{-5}\ \mathrm{nm}^{-3}$ Fig. 4c, d. The discrepancy between those answers is physically meaningful and reflects the simultaneous presence of donor ionization and acceptor neutralization. These two processes contribute similarly to net charge transfer but have opposite effects on change of disorder, leading to smaller defect densities extracted from $\Delta n_{\mathrm{dis}}$ fitting. The modelling details and the limitations of these estimations are further discussed in SI Sections 5, 6. We note, however, that both approaches should be treated as simplistic estimations of the order of magnitude of defect densities, rather than precise values.

Finally, we note that apart from defect states in hBN, we have also considered alternative mechanisms to explain our results, including interface contamination and the presence of interfacial water, which we found to be inconsistent with our results. Importantly, despite the ionization of defect states, no hysteresis is observed in any of our devices for sufficiently slow gate-voltage sweeps $\sim 1\mathrm{mV/s}$, which we typically used in high-resolution measurements near the graphene CNP. This indicates that the relevant defects equilibrate at much shorter timescales. Qualitatively, this may indicate charge exchange between closely spaced donor and acceptor defects, without direct tunnelling to graphene, as discussed further in Supplementary Section 8.

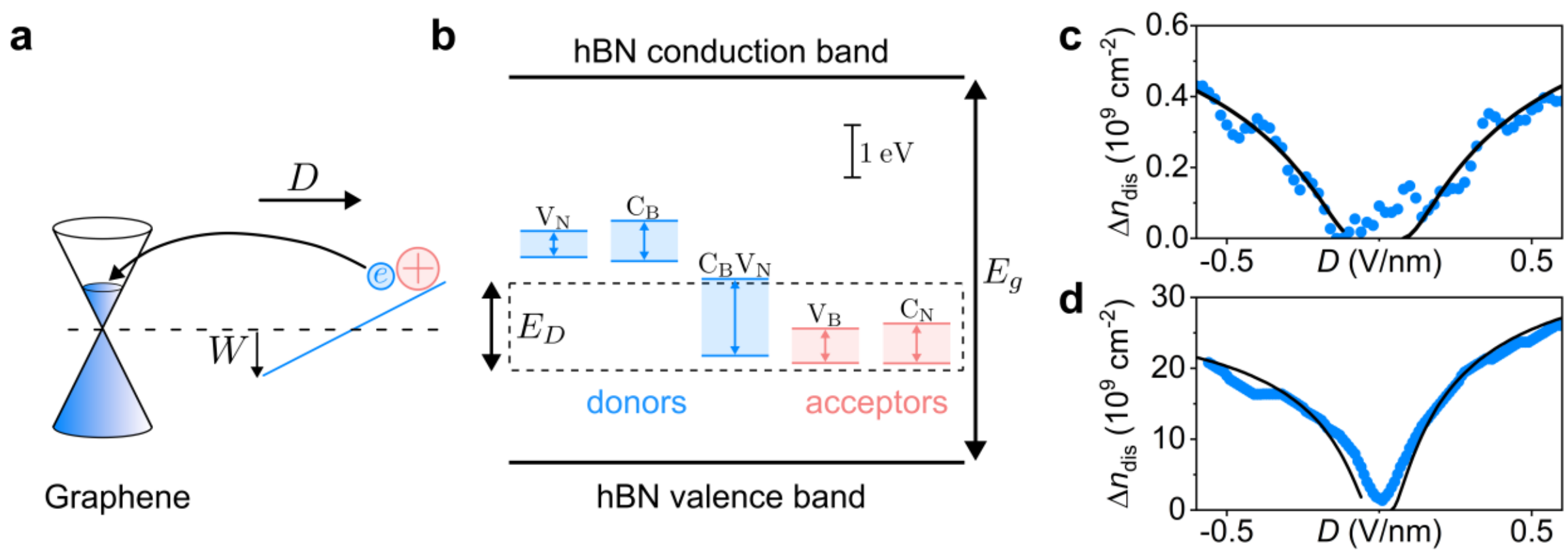

**Figure 4. Ionization of defects in hBN.** a, Schematic illustration of the spatial variation of defect energy levels under an applied displacement field relative to graphene CNP. At $D = 0$, donor states lie below the graphene CNP and therefore remain charge neutral. Applying $D$ bends the hBN bands, shifting these in-gap states above the graphene CNP, which leads to their ionization and to electron doping of graphene. b, Summary of the literature discussed in the main text. The energy diagram shows the reported position of the graphene CNP within the hBN bandgap[55–57], the range of reported acceptor-state energies[31–33] marked in red, and the range of donor-state energies[24,31–33] marked in blue. Error bars indicate the spread of values reported in different works. c,d, Fits to the relative change of device disorder estimated as $\Delta n_{\mathrm{dis}} = n_{\mathrm{HWHM}}(D) - n_{\mathrm{HWHM}}(0)$, as a function of an applied $D$ for device 1 (c) and device 5 (d). Blue symbols show the experimental data, and black lines show the model fits. See SI Section 6. In the fitting, we also introduce a cutoff $D \approx 0.05\ \mathrm{V\,nm}^{-1}$, which accounts for the finite displacement field required to start ionizing defect states located at substantial distances from graphene.

## Conclusion

In this work, we demonstrate that graphene can serve as an exceptionally sensitive charge probe for characterizing defect states in ultraclean encapsulating crystals. Our results show that even the highest-quality hBN cannot be regarded as a fully inert dielectric environment. An applied electric field induces ionization of defect states, leading to measurable charge transfer between graphene and hBN and a concurrent increase in device disorder. We emphasize that the inferred densities and types of defects are specific to the hBN source investigated here and should not be generalized to hBN produced using different conditions, equipment or growth methods. We expect, however, that the method demonstrated in our work can serve as a general approach for characterizing hBN quality across other sources.

More broadly, since the discussed defect ionization is an intrinsic property of hBN-encapsulated van der Waals heterostructures, our experiments highlight the necessity of accounting for dielectric-induced charge transfer and disorder when interpreting measurements in high-quality devices. In many systems, however, these effects are obscured by concurrent band-structure modifications, making our data particularly valuable for disentangling dielectric-induced disorder and carrier-density changes from intrinsic electronic responses at high fields in more complicated structures.

We further note that the absence of observable hysteresis under applied displacement fields in our devices indicates that defect charging remains largely reversible under typical experimental conditions. This places important constraints on the nature and dynamics of defects that have previously been invoked to explain hysteretic behavior in hBN encapsulated devices.

## Methods

### Device fabrication.

High-quality heterostructures and devices used in this work were fabricated using standard van der Waals assembly followed by nanofabrication processes[58,59]. First, monolayer graphene, thick hexagonal boron nitride (hBN, 13-37 nm), and graphite flakes (3-6 nm) were mechanically exfoliated onto oxidized Si wafers with 290 nm $SiO_2$, and flake thicknesses were identified by optical contrast and later confirmed by atomic force microscopy (AFM). Next, identified flakes were assembled into a heterostructure using a dry transfer technique based on polydimethylsiloxane (PDMS) stamps coated with a thin polycarbonate (PC) film. Exfoliated crystals were sequentially picked up, starting from a top hBN flake, which was subsequently used to collect, in order, a graphite top gate, top hBN, monolayer graphene, thin graphite contacts (for selected devices), and bottom hBN. The assembled stack was then released onto a bottom graphite gate. To avoid modification of the low-energy band structure, graphene and encapsulating hBN crystals were intentionally rotationally misaligned. To ensure high electronic quality, interfaces were formed via slow contact to squeeze away interfacial contamination, followed by high-temperature bubble removal (~130 °C) prior to final release.
Hall bar devices were defined using a standard nanofabrication workflow. First, heterostructures were characterized by high-resolution AFM to identify clean, bubble-free regions and avoid interfacial contamination in our devices. Next, electron-beam lithography (EBL) followed by $CHF_3/O_2$ reactive ion etching was used to define the Hall bar geometry. A second EBL step was then employed to pattern contact regions, followed by thermal evaporation of Cr/Au (3 nm/50 nm) to form one-dimensional graphene/graphite contacts[58]. AFM images of selected regions and optical images of final devices are shown in Supplementary Information Fig. S1.

### Electrical measurements.

Electrical transport measurements were conducted with low-frequency lock-in amplifiers from Stanford Research Systems (SRS) working at reference frequencies 9- 133 Hz. Reliability of four-point probe measurements was enhanced by using high-impedance, low-noise SR550 amplifiers. Source-drain current was adjusted to preserve the linear I-V response of the contacts over the range from 100 pA at Helium temperatures to 5 $\mu$A at elevated temperatures. Gate voltages were controlled using a high-precision source-measure unit, Keithley 2636B, with current sensitivity in the pA range. To prevent dielectric breakdown, the gate voltages were restricted to ranges in which the leakage current remained indistinguishable from the noise level of our setup, typically 10-100 pA.

## Supplementary information

**Table of contents.**

## Section 1. Details of device fabrication.

In this section, we discuss the further fabrication details of devices shown in the main text. Devices 2, 3 and 5 were assembled under ambient conditions, whereas devices 1 and 4 were fabricated inside an argon-filled glovebox. For those devices, an additional thin graphite layer (~1–2 nm) was also incorporated to enable high-quality edge contacts to graphene, and to minimize the effects of p-n junctions typically formed at the boundaries of graphite gates.

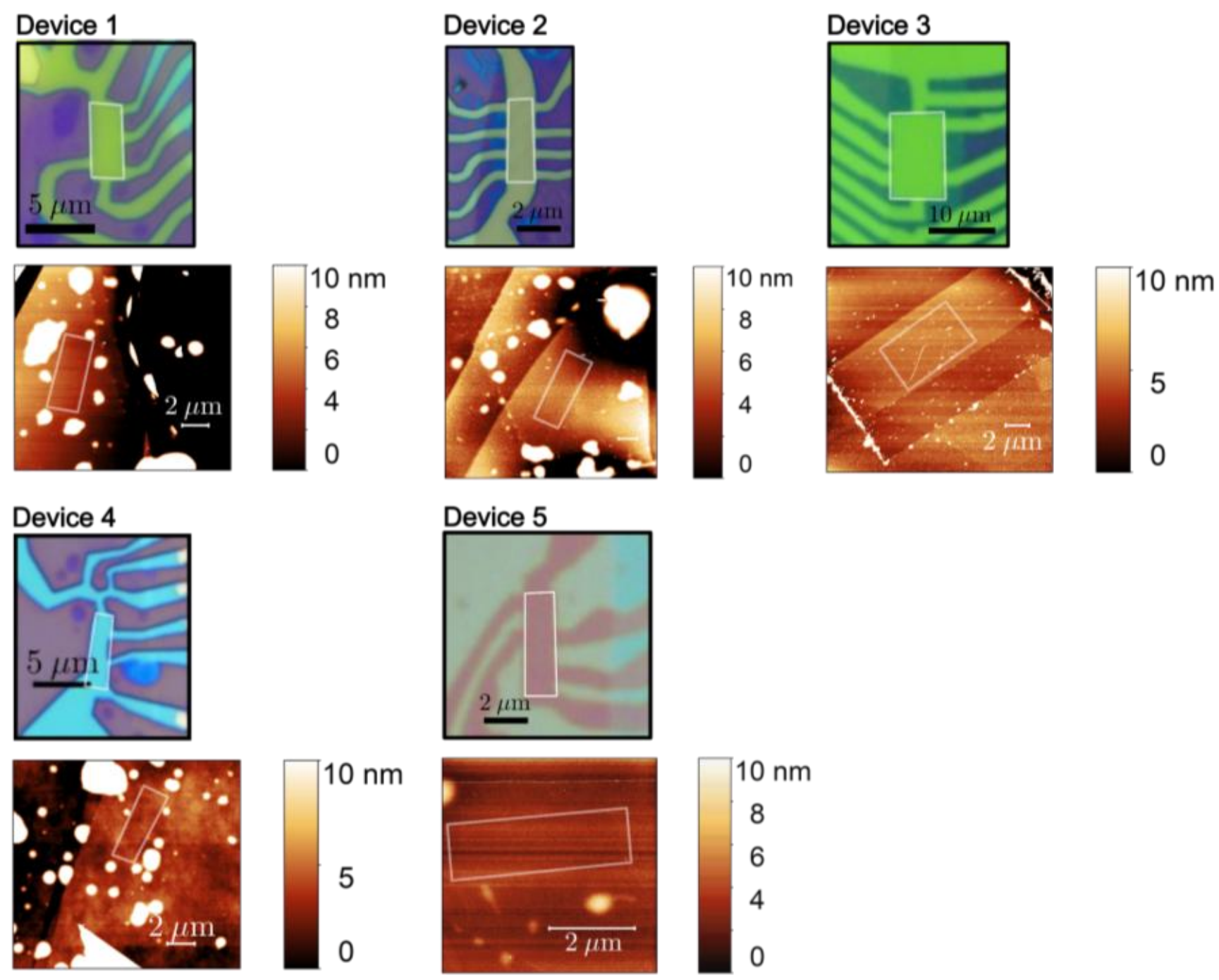

**Figure S1**. Optical micrographs and corresponding AFM scans of all double graphite-gated devices and their initial heterostructures. Rectangles in the AFM images indicate the regions selected for fabrication of the Hall bar devices.

We note that, in this work, we intentionally avoided using silicon and metallic gates. The use of a silicon back gate requires applying a gate voltage and displacement field through the $SiO_2$ layer, which hosts a significantly higher density of defects than high-quality hBN crystals. As a result, $SiO_2$ related defects are expected to dominate the electronic response under applied electric fields. Metallic top gates generally require a metal evaporation step, which can damage the encapsulating hBN and introduce additional defects. In addition, metallic films typically exhibit greater surface roughness compared to graphite, leading to increased device inhomogeneity under applied gate voltage. Finally, under applied electric fields, metallic ions can diffuse into the hBN, potentially obscuring the effects related to defect states in ultraclean hBN crystals investigated in this work.

## Section 2. Calibration of gate-induced carrier density.

In general, the gate-dependent carrier density $n$ can be written as

$$n = C_1 V_1 + C_2 V_2 + n_0, \quad (1)$$

where $C_{1/2}$ are the proportionality coefficients between corresponding top or bottom gate voltage $V_{1/2}$ and the carrier density, and $n_0$ represents the initial doping of the device at zero gate voltage. In the data analysis shown in the main text, this initial doping is typically subtracted to enable a more direct visualization of displacement field-induced effects. The extracted initial values of doping are listed in Table S1. All devices investigated in this work show substantial hole doping, which, as discussed in the main text, indicates a significant density of acceptor-like defects with energy levels well below the graphene charge neutrality point.

| Device index | 1 | 2 | 3 | 4 | 5 |
|---|---|---|---|---|---|
| $n_0$ ($10^{12}$ cm$^{-2}$) | -0.041 | -0.020 | -0.034 | -0.039 | -0.041 |

**Table S1**. Values of initial hole doping for the studied devices.

To find the calibration coefficients $C_1$ and $C_2$, we used classical Hall-effect measurements of the Hall resistance $R_H$ in the doped regime, where $R_H = B/en$ (Fig. S2a). This allows a straightforward extraction of the carrier density and, consequently, of $C_1$and $C_2$.

This approach was directly applied to analyse the data from devices 4 and 5 which exhibit relatively high disorder, as discussed further in the Supplementary Information. However, we find that this approach must be modified for devices 1-3 which are significantly cleaner. When the carrier density in these devices is tuned below $\sim 10^{10}$ cm$^{-2}$, the vanishing density of states near the Dirac point, together with the presence of a small bandgap, necessitates accounting for quantum capacitance effects. In this regime, the relation $n(V_{1/2})$ becomes nonlinear (Fig. 3a-b) and can no longer be described by the simple linear model introduced above.

In this regime, we employ a self-consistent system of equations that explicitly accounts for quantum capacitance effects[1]:

$$\begin{cases} q_1 + q + q_2 = 0, \\ \dfrac{q_1 - q - q_2}{2\varepsilon\varepsilon_0} d_1 = V_1 - \dfrac{\mu}{e}, \\ \dfrac{q_1 + q - q_2}{2\varepsilon\varepsilon_0} d_2 = -V_2 + \dfrac{\mu}{e}, \end{cases} \quad (2)$$

where $\mu = \hbar v_F \sqrt{\pi n}$ is the chemical potential of monolayer graphene, $q_1, q_2, q$ are charges on the top gate, bottom gate, and on graphene itself, respectively. The first equation represents a total neutrality condition, while the remaining two follow from Gauss's law. Accounting for quantum capacitance correction, the resulting gate-induced carrier density can be written as:

$$n = \left(\sqrt{\xi^2 + \tilde{n}} - \xi\right)^2. \quad (3)$$

Here $\tilde{n}$ is defined by geometrical capacitances and depends linearly on $V_{1/2}$: $\tilde{n} = C_1 V_1 + C_2 V_2$, while $\xi = 10\sqrt{\pi}\, \frac{\varepsilon}{8\pi} \frac{v_F}{c} \alpha^{-1} (\frac{1}{d_1} + \frac{1}{d_2})$, where $\alpha$ is the fine-structure constant; $v_F$ is the Fermi velocity of graphene;

and $c$ is the speed of light. For a typical device with two hBN dielectrics of 20 nm thicknesses $\xi \approx 0.13 \times 10^8\ \mathrm{m}^{-1}$.

In real devices, however, $\xi$ can be be further modified by electron-electron interaction induced renormalization of the Fermi velocity near the charge neutrality point[2], as well as by the presence of a bandgap. Both effects alter the low-energy spectrum and, consequently, the effective Fermi velocity in this regime. Moreover, residual charge inhomogeneity introduces some finite contribution to density of states of graphene which again should effectively suppress effects of quantum capacitance.

To extract $\xi$ for our devices, we fit the nonlinear carrier-density dependence (Fig. 3c) using Eq. (3), while fixing the geometric capacitances $C_{1/2}$, which are independently determined from Hall-effect measurements at high carrier density.

The correction parameters obtained for devices 1-3: $\xi = 0.0848 \times 10^8\ \mathrm{m}^{-1}, 0.063 \times 10^8\ \mathrm{m}^{-1},\ 0.196 \times 10^8\ \mathrm{m}^{-1}$, which is slightly smaller than the estimations obtained from the hBN thicknesses $\xi \approx 0.14 \times 10^8\ \mathrm{m}^{-1},\ 0.12 \times 10^8\ \mathrm{m}^{-1}, 0.25 \times 10^8\ \mathrm{m}^{-1}$. As a further cross-check of our calibration, we used Eq. (3) to replot the nonlinear fan diagrams (for example, Fig. 3a) as a function of carrier density. With the extracted calibration, the fan diagrams become linear, as shown in Fig. S2b, and the quantum Hall plateaus and resistance zeros in Fig. S2c-d align well with the expected filling factors. We also note that the extracted calibration is insensitive to the applied displacement field, as further illustrated in Fig. S2c-d.

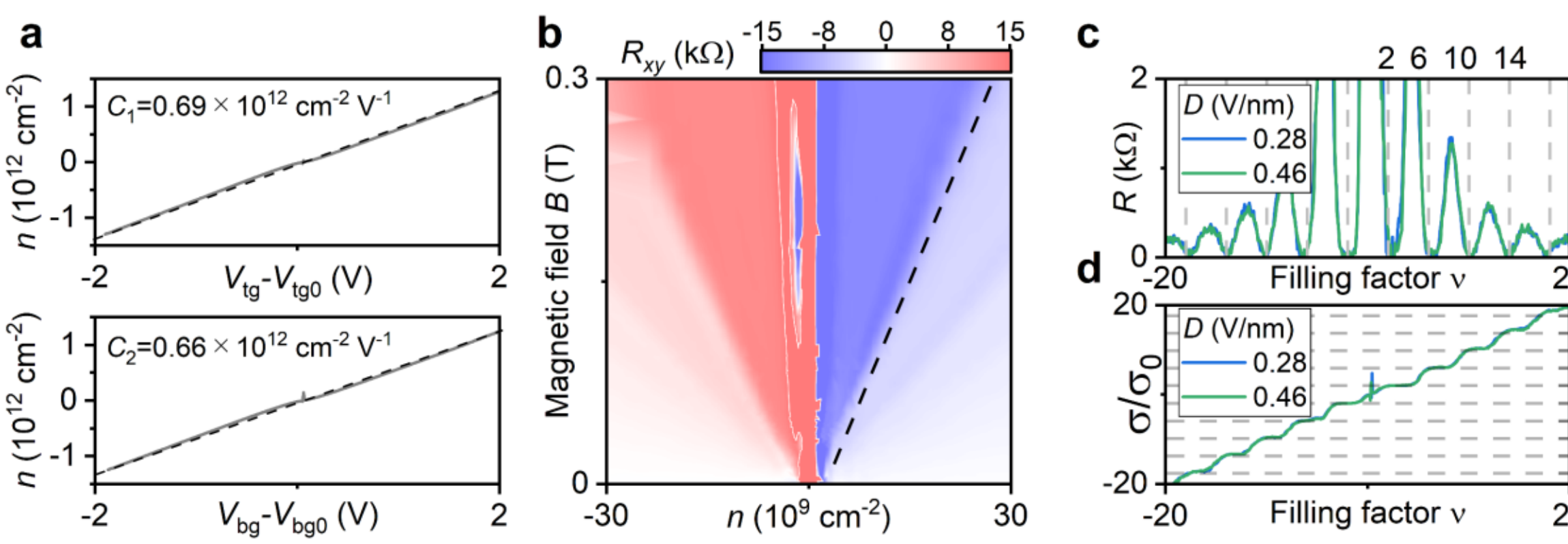

**Figure S2. a,** Fitting of calibration coefficients for top and bottom gates $C_{1/2}$. **b,** Fan diagram analogous to Fig. 3a, but plotted versus $n$, fine-tuned using Eq. (3). In these coordinates, the trajectories of Landau level gaps become straightened, as indicated with a black dashed line. **c,d,** Longitudinal resistance $R$ (c) and transverse conductivity $\sigma$ (d) in units of the conductivity quantum $\sigma_0 = \frac{e^2}{h}$ as a function of Landau filling factor $\nu$. Dashed lines correspond to the sequence of filling factors $2, 6, 10, 14, 18$. Curves for $D = 0.28\ \mathrm{V/nm}$ and $D = 0.46\ \mathrm{V/nm}$ show almost no deviation one from another, which shows the universal character of the extracted calibration coefficients $C_1, C_2$ and of the fine-tuning parameter $\xi$.

**Section 3. Uncertainties of calibration and their effect on the estimated charge transfer**

Even relatively small variations in the calibration coefficients, on the order of $0.001 \times 10^{12}\ \mathrm{cm^{-2}\ V^{-1}}$, can significantly affect the magnitude of the charge transfer discussed in the main text. See also Fig. S3. At the same time, such precision cannot be reliably achieved using the calibration procedures described above.

To further refine the calibration, we assume that in a symmetric structure (graphene is encapsulated by hBN crystals on both sides) the response to the applied displacement field should be symmetric. Based on this assumption, we introduce a fine adjustment of the calibration coefficients $C_{1/2}$ to symmetrise the position of the graphene CNP at $D = \pm 0.5\ \mathrm{V/nm}$. The required corrections to $C_{1/2}$ remain within 5% of the values obtained from the initial calibration procedure for all devices studied in this work.

We note that this assumption represents a simplification. In practice, charge transfer from the top and bottom hBN crystals may differ due to variations in defect concentration and their spatial distribution across the crystal thickness. Nevertheless, despite this uncertainty, this approach still provides a reasonable estimate of the average charge transfer under positive and negative displacement fields.

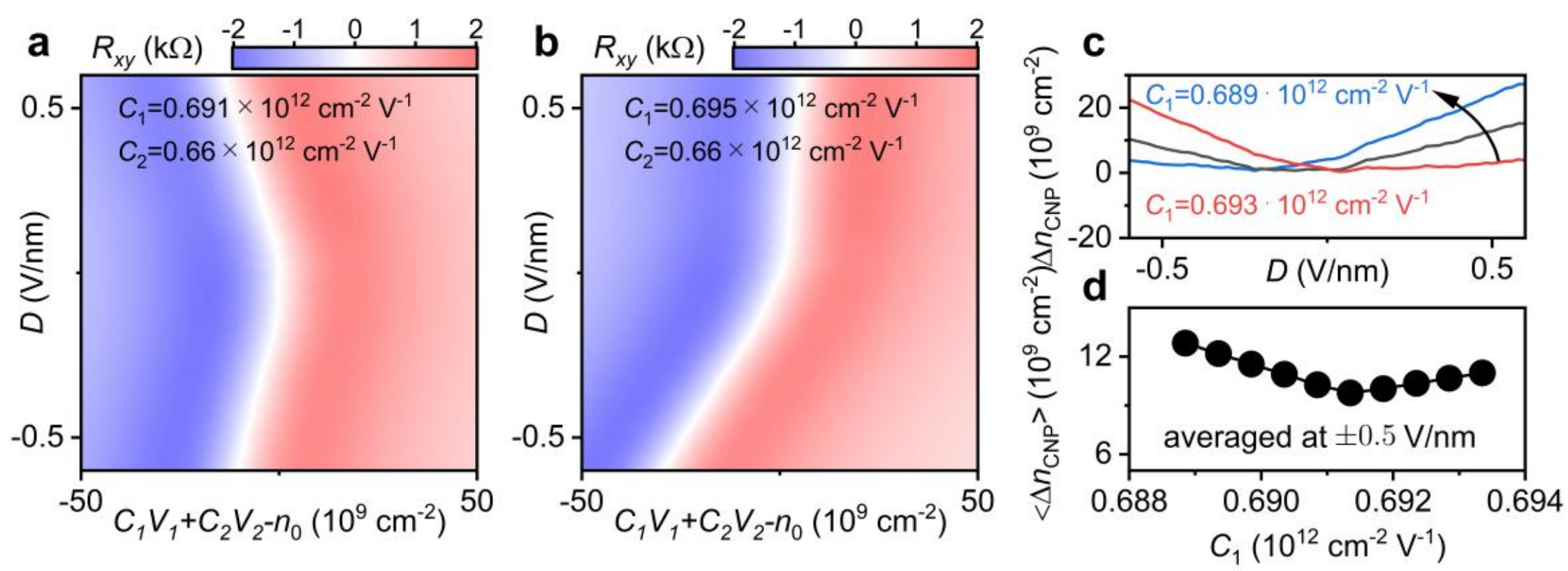

**Figure S3**. **a,b,** Double-gate maps of the Hall resistance of device 1 in $\tilde{n} - D$ coordinates for two different sets of calibration coefficients. **c,** Shift of graphene CNP obtained with slight variation of gate calibration coefficient $C_1$. **d**, CNP shift at $D = 0.5\ \mathrm{V/nm}$, averaged over positive and negative displacement fields, as a function of the calibration coefficient $C_1$. The minimum value corresponds to the $C_1$ that yields a symmetric map with respect to positive and negative $D$.

Figure S3c illustrates the charge transfer for device 1 calculated using different values of $C_1$. Despite variations in the trajectory of the curves, the charge transfer averaged at $D = \pm 0.5\ \mathrm{V\ nm^{-1}}$ consistently corresponds to electron doping of the graphene layer under the applied displacement field. The smallest values are obtained for the case of symmetric charge transfer between positive and negative $D$, indicating that our estimates provide a lower bound for the magnitude of the charge transfer in these devices.

Finally, although the magnitude of the CNP shift depends on calibration details, imperfect calibration cannot account for the observed curvature of the CNP trajectory. In the $V_1, V_2$ plane, a constant capacitance ratio $C_1/C_2$ implies a linear CNP trajectory. Even after rotating the resistance map by $\arctan(C_1/C_2)$, the CNP trajectory in our devices would remain curved for any linear adjustment of $C_1$and $C_2$. Reproducing the observed bending would require $C_1/C_2$to decrease for $V_1 < 0, V_2 > 0$ and increase for $V_1 > 0, V_2 < 0$, which is unphysical. Nonlinear capacitance corrections, including quantum capacitance, depend only on $|D|$ and not its sign, and therefore cannot explain this asymmetry. Furthermore, direct comparison of carrier densities extracted from Hall measurements, $n_{\mathrm{Hall}}$, at several $D$, with those obtained from a single calibration Fig. S2b, shows no $D$-dependent deviation within the linear regime. We therefore exclude a systematic $D$-dependence of the capacitance ratio $C_1/C_2$.

**Section 4. Additional data from measured devices.**

In the main text, we present data demonstrating the increased device inhomogeneity under an applied displacement field for devices 1 and 5. For completeness, here we show analogous measurements for devices 2-4, presented in Figs. S4-6. In all devices, the applied displacement field induces a notable charge transfer, as evidenced by the shift of the CNP in the double-gate maps shown in Fig. S4a,b, Fig. S5a,b and Fig. S6a. Simultaneously, the displacement field leads to an increase in device disorder, as shown in Fig. S4e-g, Fig. S5c-e, and Fig. S6b,c, measured using the same procedure as for devices 1 and 5 in the main text.

Although the applied displacement field produces qualitatively similar behavior in all devices, the absolute magnitude of the charge transfer, as well as the disorder at zero and finite displacement field, changes noticeably from device to device. These differences are likely due to small variations in the concentration and relative proportion of different defect states in the hBN crystals used for device fabrication.

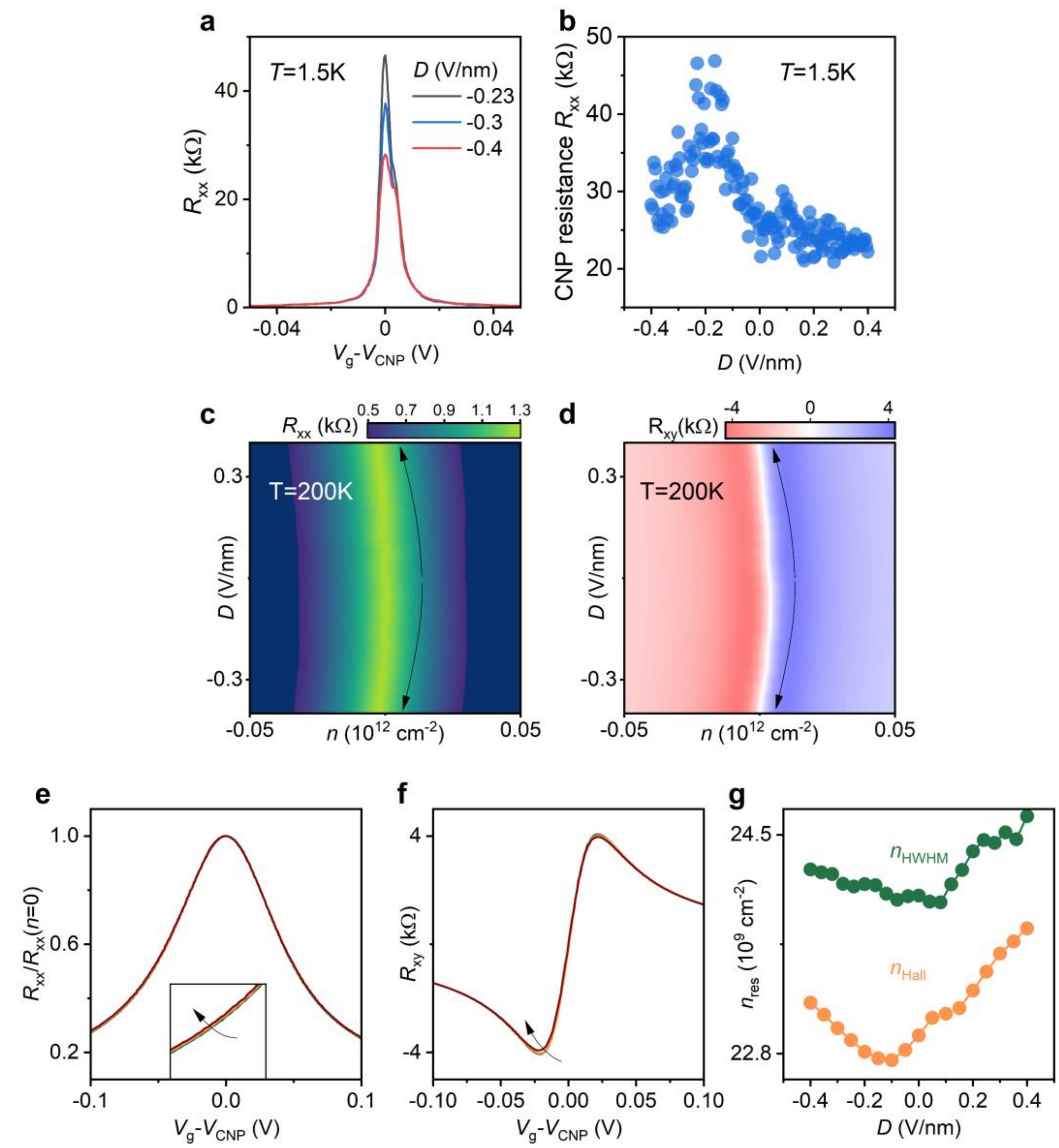

**Figure S4**. **Additional data from device 2**. **a,b,** Low-temperature $T = 1.5\mathrm{K}$ measurements analogous to Fig. 3d,e. Increasing $D$ leads to more disorder, which is reflected in suppressing high CNP resistance. **c,d,** High temperature $T = 200\mathrm{K}$ longitudinal $R_{\mathrm{xx}}$ and Hall $R_{\mathrm{xy}}$ resistance maps versus carrier density $n$ and displacement field $D$. The black arrows highlight the displacement field induced CNP shift. **e**, Normalized $R_{\mathrm{xx}}$ as a function of gate voltage relative to the CNP for selected $D$; inset highlights the broadening of the peak under applied $D$. **f**, $R_{\mathrm{xy}}$ versus gate voltage relative to the CNP for several values of $D$. $B = 150\ \mathrm{mT}$. **g,** Residual carrier density $n_{\mathrm{res}}$ as a function of $D$, extracted from the $n_{\mathrm{HWHM}}$ and $n_{\mathrm{Hall}}$.

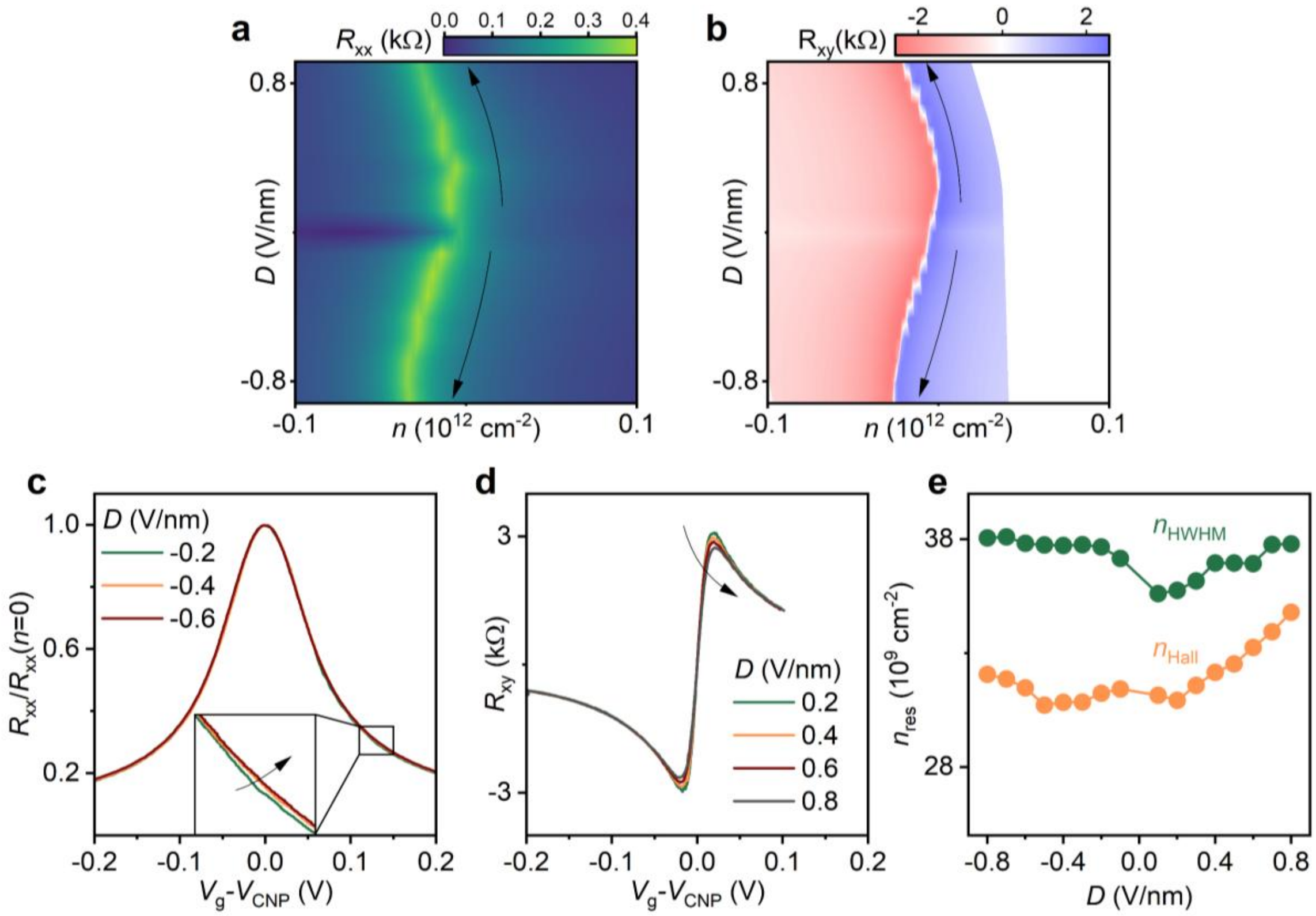

**Figure S5**. **Additional data from device 3. a,b,** Longitudinal $R_{\mathrm{xx}}$ and Hall $R_{\mathrm{xy}}$ resistance maps versus carrier density $n$ and displacement field $D$. **c**, Normalized $R_{\mathrm{xx}}$ as a function of gate voltage relative to the CNP for selected $D$. **d**, $R_{\mathrm{xy}}$ versus gate voltage relative to the CNP for several values of $D$. $B = 150\ \mathrm{mT}$. **e,** Residual carrier density $n_{\mathrm{res}}$ as a function of $D$, extracted from the $n_{\mathrm{HWHM}}$ and $n_{\mathrm{Hall}}$. Temperature $T = 200\mathrm{K}$ for all the panels.

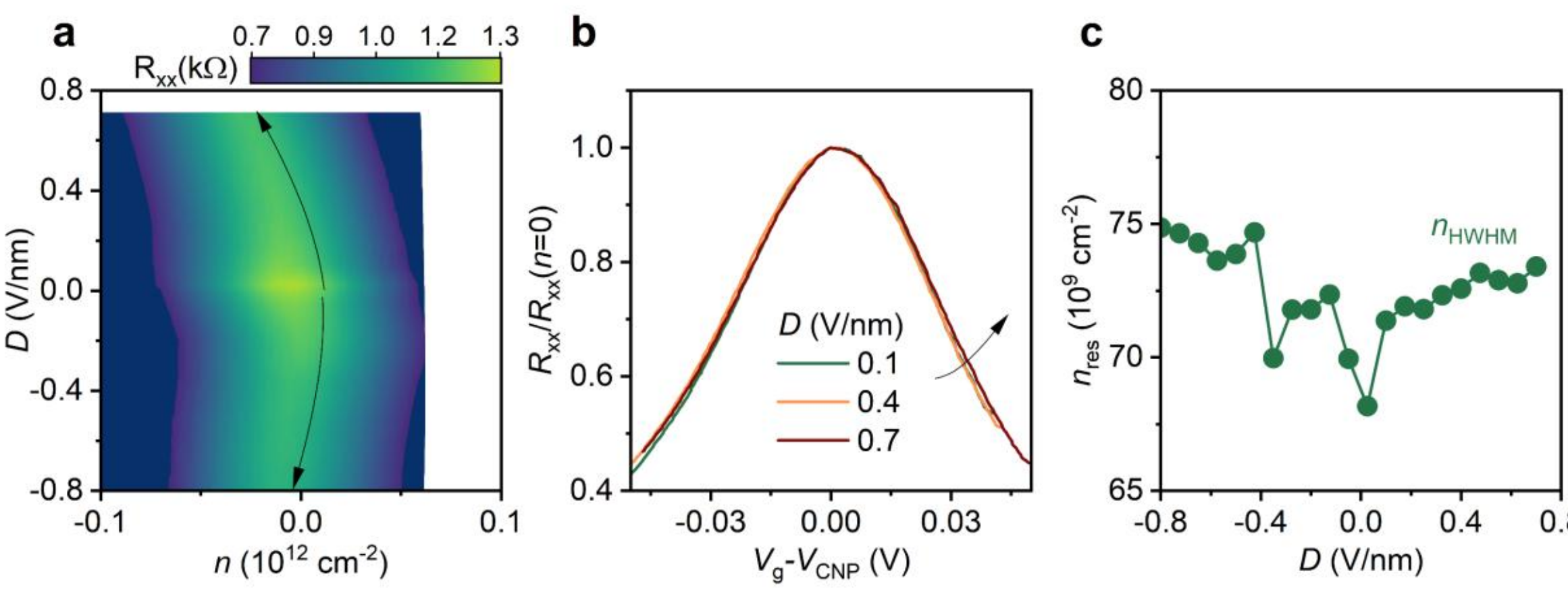

**Figure S6**. **Additional data from device 4. a**, Longitudinal $R_{\mathrm{xx}}$ resistance map versus carrier density $n$ and displacement field $D$. **b**, Normalized $R_{\mathrm{xx}}$ as a function of gate voltage relative to the CNP for selected $D$. **c**, Residual carrier density $n_{\mathrm{res}}$ as a function of $D$, extracted from the $n_{\mathrm{HWHM}}$. Temperature $T = 200\mathrm{K}$ for all the panels.

**Further measurement details of Fig. 3e.**

In the main text, for our most extensively studied Device 1, we observed $D$-dependent resistance fluctuations within the gapped state Fig. 3e, which we attribute to the discrete charging and discharging of donor and acceptor states rather than to electrical measurement noise. To support this interpretation, Fig. S7 presents two consecutive gate-voltage sweeps measured at neighboring values of $D$. A pronounced difference between the resistance traces appears near the center of the gap, whereas the smoothness of each individual trace demonstrates that measurement noise cannot account for the resistance change between them.

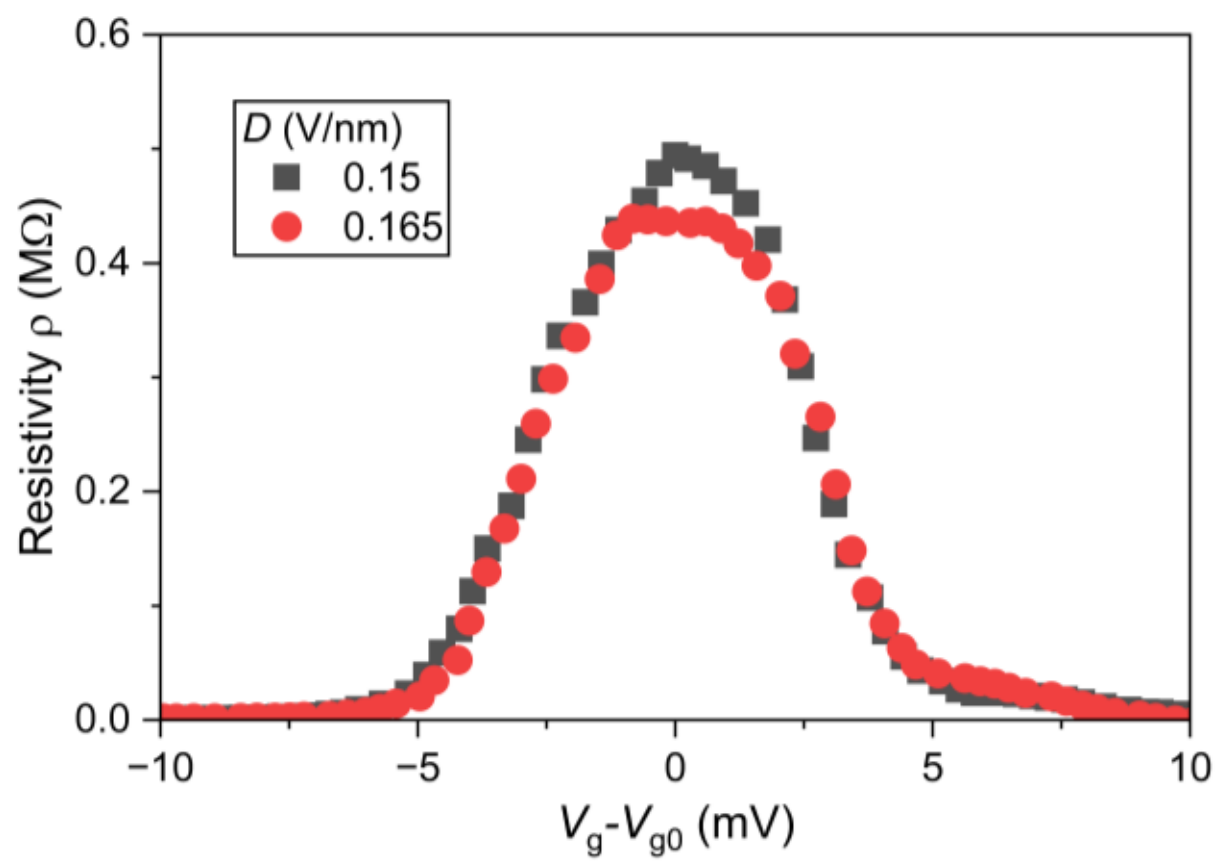

**Figure S7**. **Additional data from device 1.** $T$=1.5K. Two traces of gate-dependent resistance measured inside and outside the gapped regime for two different values of $D = 0.15$ V/nm, $0.165$ V/nm. For a small step $\Delta D = 0.015$ V/nm there is a significant change in the resistance in the gapped state, which is significantly lower than the electrical signal noise.

**Additional information about the maximum displacement field applied to our devices.**

Here we also present information about ranges of $D$ over which our samples could be measured safely without the onset of leakage currents. Although we likely did not fully explore the upper limits of $D$, this information may nevertheless be useful for studies of dielectric breakdown. Such breakdown typically requires an initial inhomogeneity to nucleate and may therefore be related to the bulk defect density of the crystal.

| Device index | 1 | 2 | 3 | 4 | 5 |
|---|---|---|---|---|---|
| Min/max $D$ (V/nm) | -1 / 1 | -0.57 / 0.68 | -0.9 / 0.8 | -0.8 / 0.85 | -0.87 / 0.9 |

**Table S2**. Lower and upper boundaries for the safe displacement field values for our devices.

Based on the data presented in Table S2, no reliable conclusion can be drawn regarding a correlation between the electric-field stability of our hBN crystals and the $D$-dependent effects observed in our measurements.

**Section 5. Estimating defect concentration based on charge transfer.**

In the main text, we discuss displacement-field-induced charge transfer between graphene and defect states in hBN. For donor and acceptor states located below the graphene Dirac point, the net effect of $D$ is n-type: donors ionize, while acceptors become neutral. In this picture, a negative (upward) displacement field drives charge exchange with the top hBN layer, whereas a positive field predominantly couples to the bottom hBN crystal.

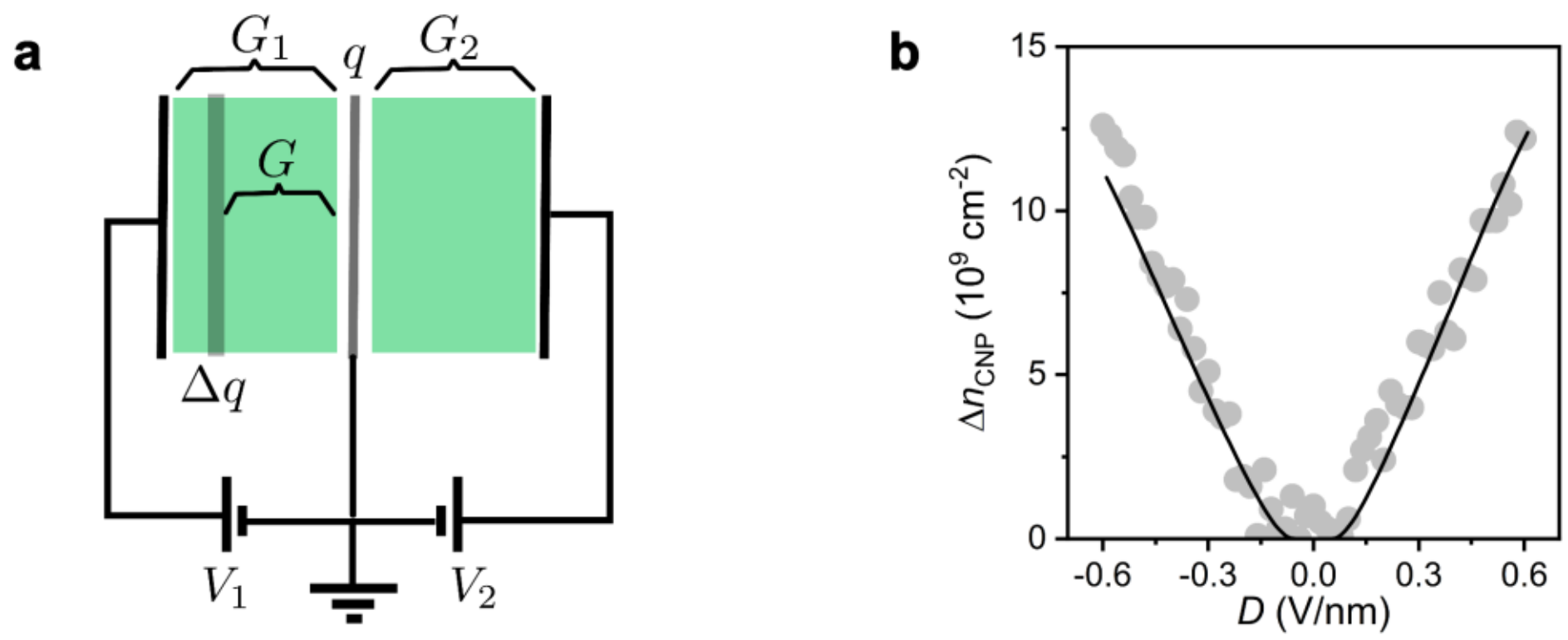

**Figure S8**. **a**, Schematic illustrating the electrostatic model implemented for analysis of CNP shifts. Charged impurities are distributed at distances $G$ from the graphene plane. **b**, The fit of the CNP shift for device 1. $G_{1/2} = 25, 26\ \mathrm{nm}$, $W_0, dW = 2.0, 1.6\ \mathrm{eV}$, and $n_i = 2.4 \times 10^{-5}\ \mathrm{nm}^{-3}$. $\varepsilon = 3.15$.

To estimate the defect-state concentration from the measured charge transfer, we employ a modified electrostatic model in which graphene exchanges charge with defects located (without loss of generality) in the top dielectric of thickness $G_1$. The electrostatic model in shown in Fig. S8a. We define $\Delta q$ as the charge transferred from graphene to the hBN layer. In this case, the electrostatic equations then become:

$$\begin{cases} q_1 + \Delta q + q + q_2 = 0, \\ \dfrac{q_1 - q - q_2}{2\varepsilon\varepsilon_0} G_1 + \dfrac{\Delta q}{2\varepsilon\varepsilon_0}(2G - G_1) = V_1 - \dfrac{\mu}{e}, \\ \dfrac{q_1 + q - q_2}{2\varepsilon\varepsilon_0} G_1 + \dfrac{\Delta q}{2\varepsilon\varepsilon_0} G_2 = -V_2 + \dfrac{\mu}{e}, \end{cases} \tag{4}$$

Solving this equation for $q$ is analogous to the initial system of equations (2), but with the substitution of $C_1V_1 + C_2V_2$ to $C_1V_1 + C_2V_2 + \Delta n_0(D)$, where $\Delta n_0(D) = \left(1 - \frac{G}{G_1}\right)\Delta q/e$ is the CNP shift at the given displacement field. We extract $\Delta n_0(D)$ from experimental data by tracing CNP shift in $\tilde{n}, D$ coordinates, which are calculated using geometrical capacitance values.

Next, $\Delta q/e$ can also be calculated by integrating the bulk density of rechargeable impurities $n_i$ over distances $G$, with integration limits set by hBN thickness $G_{1/2}$ and the $\varepsilon\, W/D$. The last one effectively describes the thickness over which defect states with the energy $W$ could be ionised/deionised (see Fig. 4a).

To describe the charge transfer, it is also necessary to account for the energy distribution of defect states. For simplicity, we assume a uniform distribution within an energy window $\Delta W$ centered at $W_0$. The resulting expression for the displacement-field-induced charge transfer is:

$$
\begin{cases}
\Delta n_0(D) = \int_{W_0-\frac{\Delta W}{2}}^{W_0+\frac{\Delta W}{2}} dW \, \frac{n_i}{\Delta W} \int_{\frac{\varepsilon W}{D}}^{G_1} dG \left(1 - \frac{G}{G_1}\right), \qquad D > \frac{\varepsilon W}{G_1}, \\
\Delta n_0(D) = 0, \qquad D < \frac{\varepsilon W}{G_1}.
\end{cases} \tag{5}
$$

We further note that the energy-offset parameters $W_0$ and $\Delta W$ can differ for the top and bottom hBN layers. An example of a fit to the experimental data using Eqs. (4) and (5) is shown in Fig. S8b. From these fits, we obtain impurity densities on the order of $10^{-5}$ nm$^{-3}$ discussed in the main text.

| Device index | 1 | 2 | 3 | 4 | 5 |
|---|---|---|---|---|---|
| $n_i$ ($10^{-5}$ nm$^{-3}$) | 2.4 | 1.6 | 6.8 | 2.6 | 1.5 |
| $W_0$ (eV) | 2.0 | 2.0 | 0.9 | 1.5 | 1.0 |
| $\Delta W$ (eV) | 1.6 | 1.7 | 0.8 | 1.3 | 0.9 |

**Table S3.** Defect densities and energy windows extracted from fits to the graphene charge neutrality point (CNP) shift under applied displacement field.

**Limitations of defect density estimations.**

To simplify our analysis, in this work we assumed a uniform distribution of donor and acceptor states both in energy within the band gap and over the crystal thickness. However, potentially one could construct non-uniform energy distributions combined with non-uniform spatial distributions across the device thickness that would produce essentially indistinguishable transport signatures. Such distributions, however, cannot be independently assessed using the transport measurements presented here.

The assumption of uniform distribution of energy offset simplifies the modeling and allows us to describe experimental data using just three fit parameters: $n_i$, $W_0$, $\Delta W$. The difference $W_0 - \Delta W$ determines the onset of ionization at low *D*, whereas $W_0$ accounts for deeper defect levels that can be ionized only at substantial distances from the graphene plane. Even within this simplified model, fitting multiple variables simultaneously limits the accuracy of the extracted parameters. Nevertheless, $W_0$ can be constrained by an approximate upper bound of 2-3 eV, corresponding to roughly half the hBN gap, which restricts the physically reasonable fitting range. With this in mind, we have been using the smallest $W_0$ which provides a reasonable fit for our data. However, we estimate the systematic relative uncertainty of the fitted parameters to be below ≈50% under the constraint that $W_0$ can be further adjusted to a higher value.

**Section 6. Estimating the concentration of defects from the change in residual carrier density at elevated temperatures**

In the main text, we demonstrate that the applied displacement field introduces a notable change in residual carrier density at elevated temperatures in all studied devices. In this section, we discuss how using $\Delta n_{\mathrm{res}}$ one can calculate the corresponding change of $\Delta n_{\mathrm{dis}}$ and estimate the corresponding concentration of defect states.

While at the helium temperatures $n_{\mathrm{res}} \approx n_{\mathrm{dis}}$, at elevated temperatures in high-quality devices $n_{\mathrm{th}} \propto T^2$ strongly contributes to the measured $n_{\mathrm{res}}$, introducing a constant background independent on the applied displacement field. To subtract this background, which is contained in $n_{\mathrm{res}}(D=0)$ we have considered the following approaches.

First, previously several theoretical works[3,4] analyzing graphene on silicon oxide have used $\Delta n_{\mathrm{res}}^2 = \Delta n_{\mathrm{th}}^2 + \Delta n_{\mathrm{dis}}^2$. However, this expression is derived under the assumption of a step-function-like electrostatic potential that creates electron–hole puddles, and is valid only when the disorder-induced potential $\delta$ is much larger than the thermal energy $kT$. In high-quality devices, in contrast, $\delta$ is expected to be only on the order of a few meV, which is much smaller than $kT$ in the temperature range $100$–$300$ K.

To overcome this limitation, we follow an approach similar to Refs.[3,4] and consider the same model potential, but instead assume the opposite limit, $\delta \ll kT$. Expanding the electron and hole densities at the charge neutrality point, shifted by $\pm\delta$ in the electron and hole puddles, to second order in $\delta/kT$, we obtain:

$$\begin{aligned} n &= n_{\mathrm{th}} + \ln 2 \frac{2}{\pi\hbar^2 v_F^2}(kT)^2 \frac{\delta}{kT} + \frac{2}{\pi\hbar^2 v_F^2}(kT)^2 \left(\frac{\delta}{kT}\right)^2, \\ p &= n_{\mathrm{th}} - \ln 2 \frac{2}{\pi\hbar^2 v_F^2}(kT)^2 \frac{\delta}{kT} + \frac{2}{\pi\hbar^2 v_F^2}(kT)^2 \left(\frac{\delta}{kT}\right)^2. \end{aligned} \tag{6}$$

Consistent with earlier definitions[3,4], $n_{\mathrm{dis}} = n - p = 2\ln 2 \frac{2}{\pi\hbar^2 v_F^2}(kT)^2 \frac{\delta}{kT}$. Then residual charge density is given as

$$n_{\mathrm{res}} = \frac{1}{2}\big(n(\delta) + n(-\delta)\big) = n_{\mathrm{th}} + \frac{2}{\pi\hbar^2 v_F^2}(kT)^2 \left(\frac{\delta}{kT}\right)^2 = n_{\mathrm{th}} + \frac{2 < V(r)^2 >}{\pi\hbar^2 v_F^2}, \tag{7}$$

where $< V(r)^2 >= \delta^2$ is a standard deviation of step-like electrostatic potential.

Next, we can also use the model of puddles with a Gaussian distribution of electrostatic potential[5]. Resulting expression, connecting $n_{\mathrm{res}}, n_{\mathrm{dis}}$, and $n_{\mathrm{th}}$ would then turn out to have its simplest form $n_{\mathrm{res}} = n_{\mathrm{dis}} + n_{\mathrm{th}}$, with $n_{\mathrm{dis}}$ expressed via standard deviation of electrostatic potential $\sqrt{< V(r)^2 >}$:

$$n_{\mathrm{dis}} = \frac{1}{2}\frac{< V(r)^2 >}{\pi\, (\hbar v_F)^2}. \tag{8}$$

To avoid overmodelling, to estimate the change of $n_{\mathrm{dis}}$ under applied displacement field we have used both step-like and Gaussian electrostatic potential models in the limit $\delta \ll kT$.

Next, the procedure used to fit the observed *D*-dependent $\Delta n_{\mathrm{HWHM}}$ is as follows. We assume that $\Delta n_{\mathrm{HWHM}}(D) \approx \Delta n_{\mathrm{res}}(D)$, and that thermal excitations are independent of the displacement field, i.e.

$\Delta n_{\mathrm{th}}(D) \approx 0$. Under these assumptions, one obtains $\Delta n_{\mathrm{HWHM}} \approx \frac{2}{\pi \hbar^2 v_F^2} \Delta\langle V(r)^2\rangle(D)$ for step-like puddles, while $\Delta n_{\mathrm{HWHM}} \approx \Delta n_{\mathrm{dis}} = \frac{1}{2\pi(\hbar v_F)^2} \Delta\langle V(r)^2\rangle(D)$ in the case of a Gaussian puddles. Since these two approaches give similar results, differing only by a numerical factor of 0.5–2, for simplicity we adopt in the following analysis

$$\Delta n_{\mathrm{HWHM}} = \frac{\Delta\langle V(r)^2\rangle(D)}{\pi(\hbar v_F)^2} \tag{9}$$

This introduces a systematic uncertainty in the extracted bulk density of rechargeable impurities, $n_i$, of up to a numerical factor of 0.5–2. However, this uncertainty does not affect the qualitative trends or the main conclusions of the study.

Next, once we know the link between the measured $\Delta n_{\mathrm{HWHM}}(D)$ and modified $< V(r)^2 > (D)$ we can estimate the concentration of ionised defect states under the applied displacement field.

The bare Coulomb potential at the graphene plane from a charge at position $z$ is (CGS units):

$$\begin{cases} \check{V}_C(q,z) = \dfrac{4\pi e^2}{\varepsilon_z q} \dfrac{\sinh(f(G_2+z)\,q)\sinh(fG_1\,q)}{\sinh(f(G_1+G_2)\,q)}, & z<0, \\ \check{V}_C(q,z) = \dfrac{4\pi e^2}{\varepsilon_z q} \dfrac{\sinh(f(G_1-z)\,q)\sinh(fG_2\,q)}{\sinh(f(G_1+G_2)\,q)}, & z>0. \end{cases} \tag{10}$$

Here $f = \sqrt{\frac{\varepsilon_x}{\varepsilon_z}}$, $\varepsilon_z = 3.15$ was found from the fit of capacitances of our devices, $\varepsilon_x = 6.7$. $G_1, G_2$ are the thicknesses of the top and bottom hBN crystals. Screening charge density of graphene is obtained from polarization $\Pi(q)$: $n_{\mathrm{screening}} = \Pi(q)V(q)$, where polarization is:

$$\Pi(q) = \frac{2k_F}{\hbar\pi v} \begin{cases} 1, & q < 2k_F \\ \dfrac{q\cos^{-1}\left(\frac{2k_F}{q}\right)}{4k_F} - \dfrac{1}{2}\sqrt{1-\dfrac{4k_F^2}{q^2}} + 1, & q \geq 2k_F \end{cases} \tag{11}$$

The potential in the graphene plane is then $\tilde{V}(q,z) = \frac{\check{V}_C(q,z)}{1+\check{V}_C(q,0)\,\Pi(q)} \approx 0.58\,\check{V}_C(q,z)$ for small doping of graphene and relatively thick hBN layers. The quantity of interest is $< V(r)^2 >$, so the inverse Fourier transform in the 2D plane over $q$ and integration over charged scatterers is done:

$$< V(r)^2 > = n_i \int_{q_{\min}}^{q_{\max}} \frac{d^2q}{(2\pi)^2} \int_{z_{\min}}^{z_{\max}} \left|\tilde{V}(q,z)\right|^2 dz\,. \tag{12}$$

Here $q_{\min}, q_{\max}$ are numerical cut-offs for $q$, defined by the size of the sample and interatomic distances as $\approx 10^6\ \mathrm{m}^{-1}, 10^9\ \mathrm{m}^{-1}$, respectively, with the result being almost independent of exact values up to the limit case $q_{\min} \to 0, q_{\max} \to \infty$. $z_{\min}, z_{\max}$ are defined by the direction of $D$ and the energy offset between the graphene Dirac point and the charge transition level of the defect. For example, for $D < 0$, ionization starts in the top hBN crystal with $z_{\min} = \frac{\varepsilon_z W}{D}$, $z_{\max} = G_1$.

We numerically integrate Eq. (12) for different values of $D$, and fit the experimental data using Eq. (9). Examples of such fits are shown in Fig. 4c,d in the main text, while a summary of the extracted impurity concentrations and energy levels is provided in Table S4.

| Device index | 1 | 2 | 3 | 4 | 5 |
|---|---|---|---|---|---|
| $n_i$ ($10^{-5}$ nm$^{-3}$) | 0.1 | 0.9 | 0.6 | 0.7 | 4.2 |
| $W$ (eV) | 0.6 | 0.7 | 0.3 | 0.3 | 0.3 |

**Table S4**. Bulk density and energy levels of rechargeable impurity states extracted for different devices from the fitting procedure described above.

From Table S4, we see a high degree of variance between different devices, which we attribute to crystal-to-crystal variability. During this study, we used hBN crystals from two different batches, produced at NIMS at least one year apart: devices 1, 2, and 4 were from the first batch, and devices 3 and 5 were from the second batch. One more possibility causing such variations in crystal quality may originate from carbon-doped domains in HPHT-synthesized hBN[6].

## Section 7. Nonlinear Hall response in the electron–hole plasma regime

As discussed in the main text, to demonstrate the change of disorder under applied displacement field we have used two estimations: $n_{\mathrm{HWHM}}$ and $n_{\mathrm{Hall}}$, but have used only $n_{\mathrm{HWHM}}$ to extract the density of ionised defects.

On the one hand, $n_{\mathrm{Hall}}$ is obtained directly from the measured resistance values, which are independent of device carrier-density calibration, which makes it tempting to use this quantity for further analysis. On the other hand, however, in our measurements we have found that close to CNP, $n_{\mathrm{Hall}}$ is sensitive to the applied magnetic field, due to a change of the Hall coefficient in this limit in our devices, as shown in Fig. S9a,b. This sensitivity is expected in the limit of Dirac plasma, for which the Hall resistance can be written as[7]:

$$R_{\mathrm{xy}} = \frac{B}{e}\frac{n-p}{(n+p)^2}\frac{1+(\mu B)^2}{1+\frac{(n-p)^2}{(n+p)^2}(\mu B)^2}. \tag{13}$$

In this expression, the magnetic field explicitly affects the shape of the $R_{\mathrm{xy}}$ peak close to CNP, leading to a field-dependent estimate of $n_{\mathrm{Hall}}$. The Hall coefficient near the CNP increases with increasing magnetic field (Fig. S9c). While, in principle, this effect can be accounted for, such an approach would rely on additional model assumptions and introduce further uncertainty. For this reason, in the present work we focus on $n_{\mathrm{HWHM}}$, which enables a more direct and robust analysis, as discussed in the previous section.

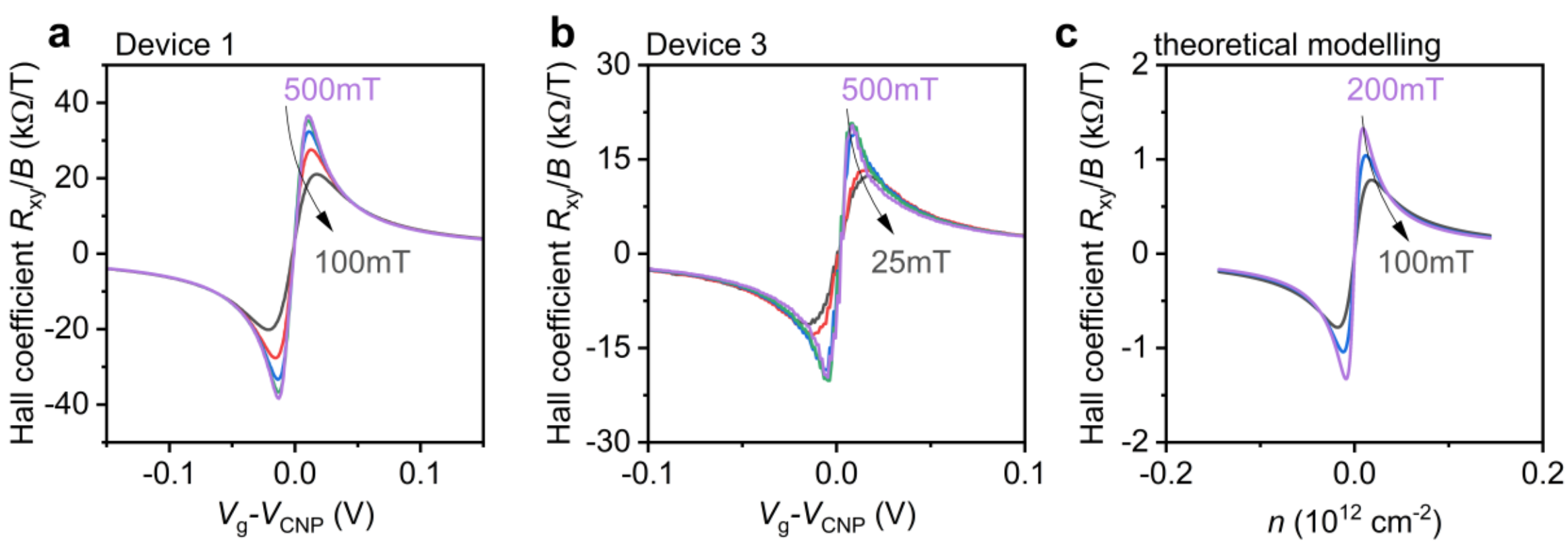

**Figure S9.** **a,b,** Hall coefficient measured in devices 1 and 3 at 200 K for different magnetic fields. **c**, Hall coefficient calculated from Eq. (13) for different magnetic fields.

**Section 8. Discussion of alternative explanations of the observed effects**

Here we consider possible alternative origins of the displacement-field-induced charge transfer and the concurrent increase in device disorder, which in the main text are attributed to ionization of defect states in hBN.

All devices were fabricated in regions verified by high-resolution AFM to be free of visible bubbles and contamination. Nonetheless, one may argue that residual nanoscale contamination (such as water molecules or small contamination clusters) was not resolved by AFM and could influence transport under an applied displacement field.

We first examine the role of interfacial water. A continuous water layer can be excluded, as graphene-water interfaces are shown to result in a strong hysteresis[8] as a function of gate voltage, which is absent in our measurements. Next, atomic-scale water islands (e.g., bilayer ice), reported on hydrophilic substrates such as mica and $BaF_2$[9,10], are not expected to form on hydrophobic hBN. Moreover, such structures would produce detectable contrast in AFM topography and would have been identified during area selection.

Next, for isolated water molecules, an applied displacement field would induce dipolar polarization, potentially enhancing disorder. However, such molecules would be present both above and below the graphene layer, leading to near cancellation of their electrostatic charging effect: dipoles oriented in opposite directions would cancel each other and produce negligible shift of the CNP. This is incompatible with the pronounced and systematic CNP shifts observed in all our devices.

Next, we also have attempted different fabrication environments for our devices. Devices 1 and 4 were assembled in an argon glovebox, where water contamination is reduced by orders of magnitude compared to ambient conditions, whereas devices 2, 3 and 5 were fabricated in air. Despite this, all devices exhibit quantitatively similar behavior, including comparable charge transfer and disorder evolution with displacement field. This strongly disfavors a water-related origin.

We next consider contamination by volatile organic compounds (VOCs) from laboratory air. Similar to water molecules, their concentration is expected to be significantly suppressed in glovebox-fabricated devices. However, no significant differences are observed between devices fabricated in inert and ambient environments. In particular, all devices also exhibit similar residual carrier densities at zero displacement field, $20 - 40 \times 10^9\ \mathrm{cm}^{-2}$. If interfacial contamination dominated the observed effects, more substantial device-to-device variability would be expected. The absence of such variability argues against this scenario.

The effects observed in our devices are consistent with previous studies in which the role of intrinsic defects was deliberately enhanced by increasing their concentration[6,11], resulting in significantly larger charge transfer and a stronger increase in device disorder than observed here.

Taken together, these considerations rule out interfacial contamination as the primary origin of the observed charge transfer and disorder increase, supporting instead an intrinsic mechanism associated with defect ionization in hBN.

To discuss our results, we also considered an additional mechanism that could explain the absence of gate hysteresis in our devices. Intuitively, at a low bulk impurity density, the characteristic distance between the graphene plane and the impurities ionized at a given *D* should increase with increasing displacement field. In this picture, the onset of long-range tunneling could lead to notable sweep-rate-dependent charge trapping in the dielectric, manifesting as gate hysteresis, which was not observed in our measurements.

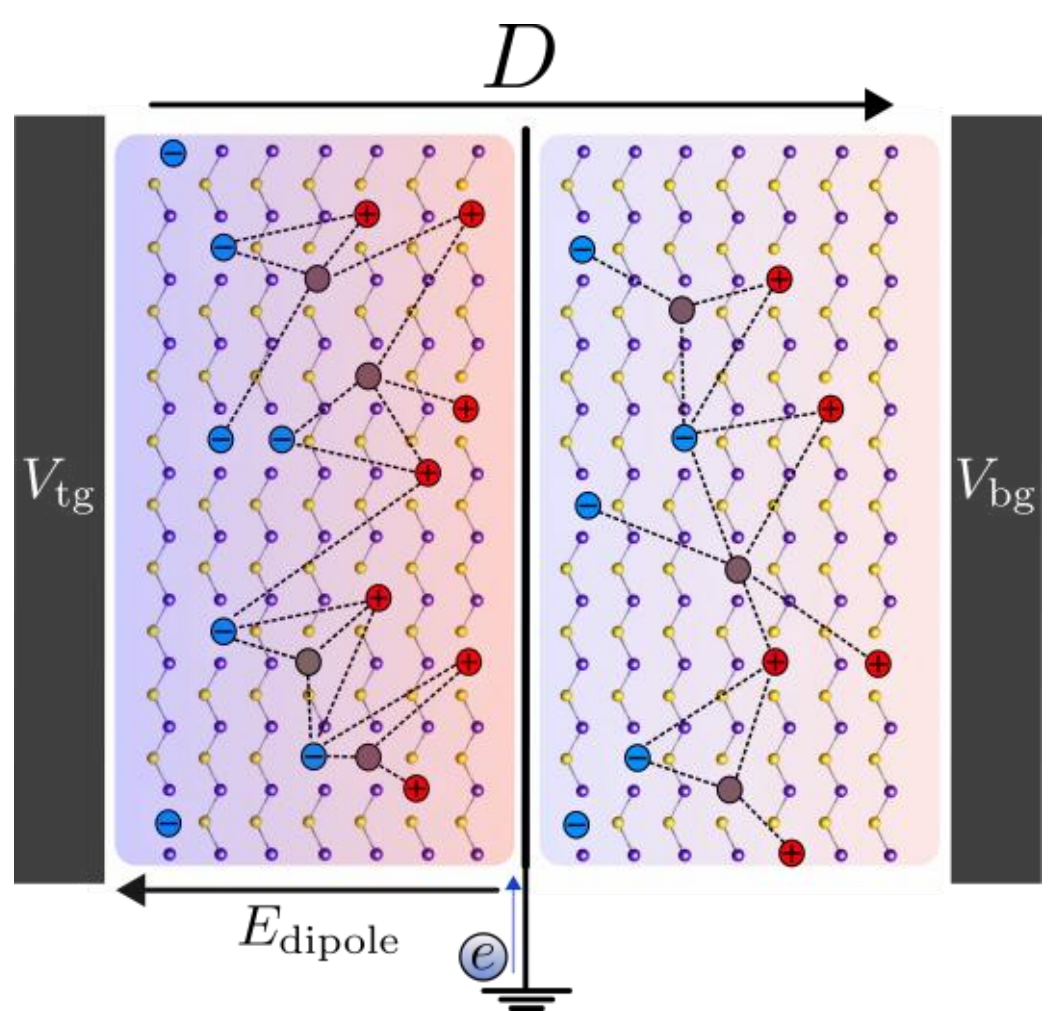

**Figure S10.** Schematic of the double-gated device with asymmetric impurity charge density redistribution. $V_{tg}, V_{bg}$ stand for top gate and bottom gate voltages, respectively. Blue, red, and grey balls represent acceptor, donor, and neutral defects; dashed lines represent possible charge tunneling pathways.

This discrepancy may be resolved by considering a modified picture of defect ionization, illustrated in Fig. S10. Even if long-range charge transfer between graphene and hBN is suppressed, defects in hBN may still be ionized through short-range charge exchange between donor and acceptor states, possibly via an intermediate tunnelling state involving a charge-neutral defect (for example, neutral acceptors, already charged acceptors changing their charge state, neutral donor-like states or other defects). Such charge transfer would create electric dipoles within the hBN. In this scenario, the additional doping observed in our devices could arise from different dipole densities in the top and bottom hBN crystals, resulting from differences in their thickness or crystal quality. The associated charge disorder would likewise originate from the spatially non-uniform distribution of these dipoles. It is also possible, however, that both mechanisms, direct ionization of defects near the graphene channel and charge exchange between donor and acceptor states located farther away, contribute simultaneously to the effects observed in our work.

**References.**